\pdfoutput=1
\documentclass[journal]{IEEEtran}
\usepackage{amsfonts}
\usepackage{amsmath}
\usepackage{upgreek}
\usepackage{amsthm}
\usepackage{graphicx}
\usepackage{color}
\usepackage{multicol}
\usepackage{amssymb}
\usepackage{subfigure}
\usepackage{bm}
\usepackage{latexsym}
\usepackage{stfloats}
\usepackage{float}
\usepackage{exscale}
\usepackage{relsize}
\usepackage{cite}
\usepackage{cases}
\usepackage{algorithm}
\usepackage{algpseudocode}
\usepackage{graphics}
\usepackage{graphicx}
\usepackage{epsfig}
\usepackage{epstopdf}
\usepackage{breqn}
\usepackage{enumerate}
\usepackage{multirow}
\usepackage{letterspace}
\usepackage[utf8]{inputenc}
\usepackage[%colorlinks=true,
linkcolor=black, %%%修改此处为你想要的颜色
anchorcolor=black, %%修改此处为你想要的颜色
citecolor=black, %%修改此处为你想要的颜色，例如修改blue为red
urlcolor=black%%设置DOI的颜色这里为蓝色
]{hyperref}
\begin{document}
	
	% paper title
	%\title{Receive Antenna Selection for a MIMO System with  Limited Feedback Aided by Channel Estimation}
	\title{Energy Minimization in RIS-Assisted UAV-Enabled Wireless Power Transfer Systems}
	
	\author{
		Hong~Ren,
		Zhenkun~Zhang,
		Zhangjie~Peng,
		Li~Li,
		and
		Cunhua~Pan
		%Jiangzhou~Wang,~\IEEEmembership{Fellow,~IEEE}
		%\thanks{Z. Peng, L. Li and Z. Zhang are with the College of Information, Mechanical and Electrical Engineering,
		% Shanghai Normal University, Shanghai 200234, China $( \text{e-mails:}$ $\text{\{pengzhangjie,}$ $\text{lilyxuan\}@shnu.edu.cn}, \text{1000479070@smail.shnu.edu.cn}
		% )$.}
	\vspace{-0.15cm}
		\thanks{
			H. Ren and C. Pan are with National Mobile Communications Research Laboratory,
			Southeast University, Nanjing, China (e-mail: hren@seu.edu.cn; cpan@seu.edu.cn).
			Z. Zhang and L. Li are with College of Information, Mechanical and Electrical Engineering,
			Shanghai Normal University, Shanghai, China (e-mail: 1000479070@smail.shnu.edu.cn; lilyxuan@shnu.edu.cn).			
			Z. Peng is with College of Information, Mechanical and Electrical Engineering, Shanghai Normal University, Shanghai, China,
			and National Mobile Communications Research Laboratory, Southeast University, Nanjing, China,
			as well as Shanghai Engineering Research Center of Intelligent Education and Bigdata, Shanghai Normal University, Shanghai, China (e-mail: pengzhangjie@shnu.edu.cn).
			\emph{ (Corresponding author: Zhangjie Peng and Zhenkun Zhang)}

			This work was supported in part by the National Key Research and Development Project under Grant 2019YFE0123600, National Natural Science Foundation of China (61701307, 62101128), Basic Research Project of Jiangsu Provincial Department of Science and Technology (BK20210205), High Level Personal Project of Jiangsu Province (JSSCBS20210105), and the open research fund of National Mobile Communications Research Laboratory, Southeast University (2018D14).
			
			Copyright (c) 20xx IEEE. Personal use of this material is permitted. However, permission to use this material for any other purposes must be obtained from the IEEE by sending a request to pubs-permissions@ieee.org.
		}
	}

	\maketitle

	\newtheorem{lemma}{Lemma}
	\theoremstyle{definition} %plain, definition or remark
	\newtheorem{theorem}{Theorem}
	\newtheorem{remark}{Remark}
	\newtheorem{corollary}{Corollary}
	\newtheorem{proposition}{Proposition}

	\renewcommand{\algorithmicrequire}{\textbf{Initialize:}}
	\renewcommand{\algorithmicensure}{\textbf{Output:}}
	
	\begin{abstract}
		Unmanned aerial vehicle (UAV)-enabled wireless power transfer (WPT) systems offer significant advantages in coverage and deployment flexibility, but suffer from endurance limitations due to the limited onboard energy.
		This paper proposes to improve the energy efficiency of UAV-enabled WPT systems with multiple ground sensors by utilizing reconfigurable intelligent surface (RIS).
		Specifically, the total energy consumption of the UAV is minimized, while meeting the energy requirement of each sensor.
		Firstly, we consider a fly-hover-broadcast (FHB) protocol, in which the UAV radiates radio frequency (RF) signals only at several hovering locations.
		The energy minimization problem is formulated to jointly optimize the UAV's trajectory, hovering time and the RIS's reflection coefficients.
		To solve this complex non-convex problem, we propose an efficient algorithm. 
		Specifically, the successive convex approximation (SCA) framework is adopted to jointly optimize the UAV's trajectory and hovering time, in which a minorization-maximization (MM) algorithm that maximizes the minimum charged energy of all sensors is provided to update the reflection coefficients.
		Then, we investigate the general scenario in which the RF signals are radiated during the flight, aiming to minimize the total energy consumption of the UAV by jointly optimizing the UAV's trajectory, flight time and the RIS's reflection coefficients.
		By applying the path discretization (PD) protocol, the optimization problem is formulated with a finite number of variables.
		A high-quality solution for this more challenging problem is obtained.
		Finally, our simulation results demonstrate the effectiveness of the proposed algorithm and the benefits of RIS in energy saving.
		
	\end{abstract}
	

	\section{Introduction}
	
	\IEEEPARstart{W}{ireless} power transfer (WPT) enabled by radio frequency (RF) transmission can continuously and steadily replenish energy for low-power devices in future Internet-of-things (IoT) networks \cite{6810996,7867826}.
	Without a wired connection for power supply, IoT devices can be deployed more flexibly and cost less to install, charge and maintain \cite{7984754}.
	Nevertheless, traditional WPT systems with fixed energy transmitter are limited in coverage and energy efficiency due to the severe signal attenuation. 
	Dense deployment of energy transmitters may lead to high construction and energy costs.

	Unmanned aerial vehicles (UAVs) operating at low altitudes are with high maneuverability and flexibility, and can be dispatched as aerial platforms for data collectors, access points, relays, etc \cite{8660516,8918497}.
	As a promising solution to improve the quality of wireless communications, the application of UAVs has received extensive attention in recent years.
	By flying close to the energy devices, UAVs are able to improve the energy harvesting efficiency of WPT systems and wireless powered communication networks (WPCNs) \cite{9234110,8982086,9273074}.
	In addition, in harsh environments such as forests, mountains, deserts and disaster areas, UAV-enabled WPT can be a more practical and cost-effective solution to extend the lifetime of low-power sensors and IoT devices than terrestrial infrastructures.
	However, when deploying UAVs in urban areas to track the cars or monitor the dynamic traffic status, the communication link between the UAVs and the ground devices may suffer from the blockages due to the high dense trees, buildings, advertising board, etc. 
	This will result in low energy harvesting efficiency. 
	
	As a promising technology for the future wireless communication networks, reconfigurable intelligent surface (RIS) has attracted extensive research attention recently.
	It improves the quality of wireless communications by reconfiguring the radio propagation environment \cite{9140329,wu2019towards}.
	With the aid of RISs, the performance of the existing wireless communication systems can be improved significantly \cite{9366346,9180053,8811733}.
	Furthermore, the integration of RIS into various communication systems has been proposed, for multicell MIMO communications \cite{9090356}, full-duplex cellular communications \cite{9318531}, mobile edge computing \cite{bai2019latency}, and simultaneous wireless information and power transfer (SWIPT) \cite{9110849}.
	An RIS is a low-power device with no RF chain, which is a programmable metamaterial planar surface that consists of a large number of passive reflecting elements \cite{tie2017information}.
	Each element independently reflects the incident signal while inducing a digitally controlled phase shift \cite{liu2019intelligent}.
	Thanks to the developments in metamaterials, the phase shifts can be reconfigured in real time \cite{9475160}.
	When they are properly adjusted, an RIS can provide substantial passive beamforming gains.
	Moreover, the signal power at a given receiver can be enhanced by constructively superimposing the reflected and direct signals.
	It is a cost-efficient solution to integrate the RIS into UAV-enabled communication systems \cite{2021arXiv210307151Y,2020arXiv201204775S,2021arXiv210900876Y},
	and has demonstrated considerable performance enhancement in a variety of scenarios, such as data collection \cite{2021arXiv210317162A}, mobile relaying \cite{9206550} and symbiotic radio systems \cite{9400768}.

	Inspired by these advantages, the integration of RIS in UAV communications with WPT has been explored in various scenarios.
	In \cite{datadownlinksensor}, the authors proposed a device called the power and data beacon (PDB) to enable the downlink energy and information delivery in systems with a vast number of IoT nodes, and considered the deployment of RIS to assist the information transmission from UAV to PDBs.
	Based on the technologies of RIS and UAV-mounted cell-free massive MIMO, the authors in \cite{9524984} proposed a novel framework for time-division energy harvesting of IoT devices.
	Considering a UAV-enabled WPCN in which IoT devices upload data using the harvested energy with the assistance of RIS, the authors in \cite{2021arXiv210802889K} utilized deep reinforcement learning to maximize the total network rate. 
	In \cite{2021arXiv210711016L}, the sum-rate maximization problem was investigated for an RIS-empowered UAV SWIPT scheme, in which a single-antenna UAV simultaneously transmits signals to multiple users, each of which harvests energy with a power splitter.
	However, as an important challenge hindering the practical application of UAV, the energy consumption has not been considered in the above RIS-assisted WPT scenarios.
	
	In this paper, we consider an RIS-assisted UAV-enabled WPT system, where the UAV is dispatched to provide the required power for multiple ground sensors with minimal energy consumption assisted by an RIS.
	To maximize the harvested energy, unlike time-division energy transfer scenarios \cite{datadownlinksensor,9524984}, all sensors should always receive the RF signals in our considered scenarios.
	In this case, it is not optimal to align the phase of the signal transmitted through one UAV-RIS-sensor link with that through the corresponding UAV-sensor direct link.
	Although the popular semidefinite relaxation (SDR)-based method can be adopted to solve the problems, it has high computational complexity.
	To address this issue, efficient algorithms are developed for solving the formulated problems.
	Specifically, we summarize the main contributions of our work as follows:
	\begin{enumerate}
		\item 
		To the best of our knowledge, this is the first attempt to explore the benefits of reducing the system energy consumption by deploying an RIS in a UAV-enabled WPT system.
		Specifically, we consider a WPT system with multiple sensors, and propose to minimize the total energy consumption of the UAV while ensuring sufficient power supply for all sensors.
%		By applying the PD protocol, the energy minimization problem is formulated with finite number of variables.
%		The energy minimization problem in a intuitive scenario and that in the general scenario are investigated.
		Due to the continuous reception of RF signals by all sensors, the variables in the energy constraints of the energy minimization problems are coupled.
		Moreover, the complicated non-convex expression of energy consumption and the unit-modulus constraints aggravate the difficulty of solving these problems.
		
		\item
		Under the fly-hover-broadcast (FHB) protocol, we formulate the energy minimization problem for a special scenario and propose an efficient iterative algorithm to solve it.
		The FHB protocol requires the UAV to visit several hovering locations in sequence and to radiate RF signals only at those locations, which can also serve as the performance lower bound of RIS-assisted UAV-enabled WPT systems.
		We design an efficient iterative algorithm based on the successive convex approximation (SCA) framework and the minorization-maximization (MM) method. 
		Specifically, in the inner iteration, an MM algorithm is used to optimize the reflection coefficients of the RIS by maximizing the minimum charged energy of all sensors.
		
		\item
		We extend the proposed method to solve the energy minimization problem in the general scenario.
		Specifically, we apply a \emph{path discretization} (PD) protocol \cite{8663615} that enables the UAV to radiate RF signals during its flight, and formulate the problem with a finite number of variables.
		Note that the propulsion energy of UAV accounts for the vast majority of the total energy consumption.
		The PD protocol brings significant energy saving by reducing the total flight time and providing more freedom for the planning of trajectory and flying speed.  % \cite{2021arXiv210900876Y}
		
		\item
		Simulation results validate the energy efficiency improvement brought by an RIS over traditional UAV-enabled WPT systems, and verify the performance superiority of PD protocol over FHB protocol.
		Simulation results also validate the performance advantages of the proposed
		algorithms over the SDR-based benchmark.
		
	\end{enumerate}
	
	The remainder of this paper is organized as follows.
	Section \ref{SYSTEM_MODEL} describes the system model.
	In Section \ref{FLY-HOVER-BROADCAST PROTOCOL}, we formulate the energy minimization problem under the FHB protocol, and develop an efficient algorithm.
	In Section \ref{PATH DISCRETIZATION PROTOCOL}, the proposed method is extended to the general scenario under the PD protocol.
	Extensive simulation results are shown in Section \ref{SIMULATION RESULTS}.
	Finally, conclusions are drawn in Section \ref{CONCLUSION}.
	
	\emph{Notations}:
	Boldface lowercase and uppercase letters respectively denote vectors and matrices.
	$j$ and ${\mathbb C}$ denote the imaginary unit $ \sqrt {{\rm{ - 1}}} $ and the complex field, respectively.
	The conjugate, transpose, Hermitian and trace of a matrix ${\bf X}$ are denoted by ${\bf X}^{*}$, ${\bf X}^{\rm T}$, ${\bf X}^{\rm H}$ and $ {\mathop{\rm Tr}} \left( {\bf{X}} \right) $, respectively.
	For a vector $ {\mathbf{x}} $, $\left\| {\mathbf{x}} \right\|_1$ and $\left\| {\mathbf{x}} \right\|_2$ are its $l_1$- and $l_2$-norm, respectively.
	$\left|x\right|$, ${\rm Re}\left\{x\right\}$, ${\mathbb E}\left\{x\right\}$ and $\angle\left(x\right)$ denote the absolute value, real part, expectation, and angle of a scalar $x$, respectively.

	\begin{figure}
		\centering
		\includegraphics[width=0.8\linewidth]{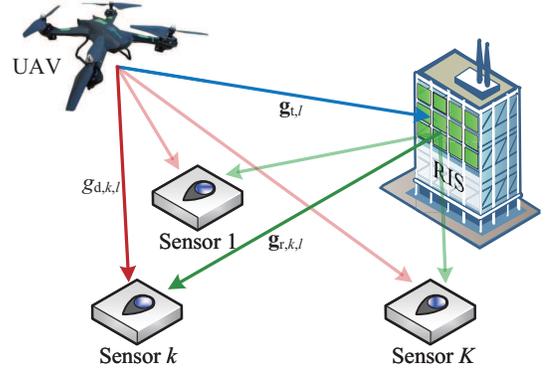}
		\caption{System model of the RIS-assisted UAV-enabled WPT system.}
		\label{Figsysmodel}
	\end{figure}

	\section{System Model}\label{SYSTEM_MODEL}
%	\subsection{System Model}
	Consider a WPT system as shown in Fig. \ref{Figsysmodel}, where one rotary-wing UAV with a single transmit antenna is dispatched as a mobile power transmitter to serve $K$ ground sensors with the aid of an RIS.
	The UAV is assumed to fly at a fixed altitude $H_{\mathrm{U}}$ with the maximum speed of $V_{\max}$ and a fixed radiated power $P_{\mathrm{t}}$.
	Each sensor is equipped with a single receive antenna and a rechargeable battery.
	The required energy for Sensor $k$ is denoted by ${E_{k}^{\mathrm{req}}}$.
	The positions of all sensors are \textit{apriori} known by the UAV.
	During each UAV's flight, all sensors continuously charge themselves by harvesting energy from the radiated RF signals, which is enhanced by the RIS.
	To facilitate the expression in the following discussion, we express the x and y coordinates of the devices by the real and imaginary parts of a complex number, respectively.
	For instance, the coordinates of Sensor $k$ is denoted by $q_{\mathrm{S},k}=q_{\mathrm{S},k}^{x}+jq_{\mathrm{S},k}^{y}$.
	Additionally, the RIS is deployed at $q_{\mathrm{R}}=q_{\mathrm{R}}^{x}+jq_{\mathrm{R}}^{y}$ with altitude $H_{\mathrm{R}}$.

%	\subsection{Energy Consumption and Power Transmission Model}
	For a rotary-wing UAV with speed $ v $, the propulsion power can be modeled as \cite[Eq. (12)]{8663615}
	\begin{align}\label{P_p}
		P_{\mathrm{p}}\left( v \right)&=P_0\left( 1+\frac{3v^{2}}{U_{\mathrm{tip}}^{2}} \right) +P_{\mathrm{i}}\left( \sqrt{1+\frac{v^{4}}{4v_{0}^{4}}}-\frac{v^{2}}{2v_{0}^{2}} \right) ^{\frac{1}{2}} \notag \\
		&\quad+\frac{1}{2}d_0\rho sAv^{3}.
	\end{align}
	The physical meanings of the parameters in \eqref{P_p} are explained in Table \ref{Table_P_p}, and the definitions of which are detailed in the appendix of \cite{8663615}.

	\begin{table}[tb]
		\begin{center}
			\caption{Notations and Terminologies of Propulsion Power}
			\label{Table_P_p}
			\begin{tabular}{c l} 
				\hline\hline\noalign{\smallskip}
				\textbf{Notation} & \textbf{Physical meaning} \\
				\hline\noalign{\smallskip}
				$P_0$ & Blade profile power in hovering status\\
				$P_{\mathrm{i}}$ & Induced power in hovering status\\
				$U_{\mathrm{tip}}$ & Tip speed of the rotor blade \\
				$v_0$ & Mean rotor induced velocity in hover \\
				$d_0$ & Fuselage drag ratio \\
				$\rho$ & Air density \\
				$s$ & Rotor solidity \\
				$A$ & Rotor disc area \\
				\noalign{\smallskip}\hline
			\end{tabular}
		\end{center}
	\end{table}
	
	For simplicity, we assume that the RIS is equipped with a uniform linear array consisting of $M$ passive reflecting elements.
	Denote by $\theta _{m,q}$ the phase shift of the $m$th reflecting element of the RIS when the UAV is at location $q=q^{x}+jq^{y}$.
	The reflection coefficient vector is defined as ${{\bm{\phi}}}_q=\left[ e^{j\theta _{1,q}},...,e^{j\theta _{M,q}} \right] ^{\mathrm{T}}$.
	%	The constraints on $\left\{ {{\bm{\phi}} }_q \right\}$ is given by
	The elements in $ {{\bm{\phi}} }_q $ satisfy
	\begin{equation}\label{constraint_theta}
		\theta _{m,q}\in\left[ 0,2\pi \right) ,\; m = 1, \dots,M.
	\end{equation}
	When the UAV is at location $q$, the baseband channels from the UAV to Sensor $k$, from the UAV to the RIS, and from the RIS to Sensor $k$ are denoted by
	$g_{\mathrm{d},k,q}\in \mathbb{C} ^{1\times 1}$, $\mathbf{g}_{\mathrm{t},q}\in \mathbb{C} ^{M\times 1}$ and $\mathbf{g}_{\mathrm{r},k,q}\in \mathbb{C} ^{M\times 1}$, respectively.
	The Rician fading is assumed for all channels.
	The channels related to the RIS are modeled as
	\begin{align}
		\mathbf{g}_{\mathrm{t},q}&=\sqrt{\beta _{\mathrm{t},q}}\left( \sqrt{\frac{\kappa _{\mathrm{t}}}{\kappa _{\mathrm{t}}+1}}\mathbf{g}_{\mathrm{t},q}^{\mathrm{LoS}}+\sqrt{\frac{1}{\kappa _{\mathrm{t}}+1}}\mathbf{g}_{\mathrm{t},q}^{\mathrm{NLoS}} \right), \label{g_t}\\
		\mathbf{g}_{\mathrm{r},k,q}&=\sqrt{\beta _{\mathrm{r},k}}\left( \sqrt{\frac{\kappa _{\mathrm{r}}}{\kappa _{\mathrm{r}}+1}}\mathbf{g}_{\mathrm{r},k}^{\mathrm{LoS}}+\sqrt{\frac{1}{\kappa _{\mathrm{r}}+1}}\mathbf{g}_{\mathrm{r},k,q}^{\mathrm{NLoS}} \right),\label{g_r}
	\end{align}
	where $\beta _{\mathrm{t},q}$ and $\beta _{\mathrm{r},k}$ represent the large-scale fading coefficients,
	${\kappa _{\mathrm{t}}}$ and ${\kappa _{\mathrm{r}}}$ denote the Rician factors,
	$\mathbf{g}_{\mathrm{t},q}^{\mathrm{LoS}}$ and $\mathbf{g}_{\mathrm{r},k}^{\mathrm{LoS}}$ denote the deterministic LoS channels, 
	$\mathbf{g}_{\mathrm{t},q}^{\mathrm{NLoS}}$ and $\mathbf{g}_{\mathrm{r},k,q}^{\mathrm{NLoS}}$ are the small-scale channel fading.
	Each element of $\mathbf{g}_{\mathrm{t},q}^{\mathrm{NLoS}}$ and $\mathbf{g}_{\mathrm{r},k,q}^{\mathrm{NLoS}}$ independently follows the circularly symmetric complex Gaussian (CGCS) distribution with zero mean and unit variance.
	%	In addition, all elements of NLoS channel components $\mathbf{g}_{\mathrm{t},q}^{\mathrm{NLoS}}$ and $\mathbf{g}_{\mathrm{r},k,q}^{\mathrm{NLoS}}$ are independently and identically distributed with zero mean and unit variance.
%	Similar definitions are used for $\beta _{\mathrm{r},k}$, ${\kappa _{\mathrm{r}}}$, $\mathbf{g}_{\mathrm{r},k}^{\mathrm{LoS}}$ and $\mathbf{g}_{\mathrm{r},k,q}^{\mathrm{NLoS}}$.
	The expressions of $\beta _{\mathrm{t},q}$ and $\beta _{\mathrm{r},k}$ are respectively given by
	\begin{align}
		\beta _{\mathrm{t},q}&=\frac{\beta _0}{\left( \left| q-q_{\mathrm{R}} \right|^2+\left( H_{\mathrm{U}}-H_{\mathrm{R}} \right) ^2 \right) ^{\alpha _{\mathrm{t}}/2}}\triangleq \frac{\beta _0}{d_{\mathrm{t},q}^{\alpha _{\mathrm{t}}}}, \\
		\beta _{\mathrm{r},k}&=\frac{\beta _0}{\left( \left| q_{\mathrm{S},k}-q_{\mathrm{R}} \right|^2+H_{\mathrm{R}}^{2} \right) ^{\alpha _{\mathrm{r}}/2}}\triangleq \frac{\beta _0}{d_{\mathrm{r},k}^{\alpha _{\mathrm{r}}}},
	\end{align}
	where $\beta _0$ is the channel gain at the reference distance of 1 m, 
	$d_{\mathrm{t},q}$ is the distance between the UAV and the RIS, $d_{\mathrm{r},k}$ is the distance between the RIS and Sensor $k$,
	and the corresponding path loss coefficients are denoted by $ \alpha _{\mathrm{t}} $ and $ \alpha _{\mathrm{r}} $, respectively.
	Based on the assumption of uniform linear array, the LoS components of the RIS-related channels $\mathbf{g}_{\mathrm{t},q}^{\mathrm{LoS}}$ and $\mathbf{g}_{\mathrm{r},k}^{\mathrm{LoS}}$ are respectively modeled as
	\begin{align}
		\mathbf{g}_{\mathrm{t},q}^{\mathrm{LoS}}&=e^{-j\frac{2\pi d_{\mathrm{t},q}}{\lambda}}\left[ 1,e^{-j\frac{2\pi d}{\lambda}\cos \omega _{\mathrm{t},q}},...,e^{-j\frac{2\pi (M-1)d}{\lambda}\cos \omega _{\mathrm{t},q}} \right] ^{\mathrm{T}}, \\
		\mathbf{g}_{\mathrm{r},k}^{\mathrm{LoS}}&=e^{-j\frac{2\pi d_{\mathrm{r},k}}{\lambda}}\left[ 1,e^{-j\frac{2\pi d}{\lambda}\cos \omega _{\mathrm{r},k}},...,e^{-j\frac{2\pi (M-1)d}{\lambda}\cos \omega _{\mathrm{r},k}} \right] ^{\mathrm{T}},
	\end{align}
	where $\lambda$ and $d$ denote the wavelength and element spacing of the RIS, respectively,
	and $\cos \omega _{\mathrm{t},q}=\frac{q_{\mathrm{R}}^{x}-q^{x}}{d_{\mathrm{t},q}}$ and $\cos \omega _{\mathrm{r},k}=\frac{q_{\mathrm{S},k}^{x}-q^{x}}{d_{\mathrm{r},k}}$ represent the cosine of angle-of-arrival (AoA) and angle-of-departure (AoD), respectively.
	Similarly, the direct link from the UAV to Sensor $k$ is modeled as
	\begin{equation}\label{g_d}
		g_{\mathrm{d},k,q}=\sqrt{\beta _{\mathrm{d},k,q}}\left( \sqrt{\frac{\kappa _{\mathrm{d}}}{\kappa _{\mathrm{d}}+1}}g_{\mathrm{d},k,q}^{\mathrm{LoS}}+\sqrt{\frac{1}{\kappa _{\mathrm{d}}+1}}g_{\mathrm{d},k,q}^{\mathrm{NLoS}} \right)\!,\!
	\end{equation}
	where $\beta _{\mathrm{d},k,q}=\frac{\beta _0}{d_{\mathrm{d},k,q}^{\alpha _{\mathrm{d}}}}$ is the large-scale fading coefficient,
	$d_{\mathrm{d},k,q}=\sqrt{\left| q-q_{\mathrm{S},k} \right|^2+H_{\mathrm{U}}^{2}}$ is the distance between the UAV and Sensor $k$,
	$g_{\mathrm{d},k,q}^{\mathrm{LoS}}=\exp \left( -j\frac{2\pi}{\lambda}d_{\mathrm{d},k,q} \right) $ is the LoS channel component,
	and the NLoS channel component $ g_{\mathrm{d},k,q}^{\mathrm{NLoS}} $ follows zero-mean and unit-variance CGCS distribution.
	Let us introduce a diagonal reflection coefficient matrix $\mathbf{\Phi }_q={\mathrm{diag}}\left( {\bm{\phi}}_q \right) $.
	Then, we can express the equivalent channel spanning from the UAV to Sensor $k$ as $\tilde g_{k,q}= g_{\mathrm{d},k,q}+\mathbf{g}_{\mathrm{r},k,q}^{\mathrm{H}}\mathbf{\Phi }_q\mathbf{g}_{\mathrm{t},q}$.
	Accordingly, the received power at Sensor $k$ from the RF signal radiated at location $q$ is given by
	\begin{equation}\label{P_k,l_origin}
		P_{k,q}=P_{\mathrm{t}}\left| g_{\mathrm{d},k,q}+\mathbf{g}_{\mathrm{r},k,q}^{\mathrm{H}}\mathbf{\Phi }_q\mathbf{g}_{\mathrm{t},q} \right|^2.
	\end{equation}
	Without loss of generality, we assume the same energy conversion efficiency $\eta $ for all sensors.

	Note that in practice, although channel estimation methods for RIS-assisted networks have been proposed in the existing literature \cite{9665300}, it is impossible to estimate the small scale channel fading before the flight.
	Hence, we design the system based on the long-term channel state information.  	%	Therefore, the statistical channel state information (CSI) have to be adopted.
	To this end, the following theorem provides the expectation of the received power at each sensor $ \hat{P}_{k,q} =\mathbb{E} \left\{ P_{k,q} \right\}$.
	\begin{theorem}\label{theorem_hatP}
		The expected received power $ \hat{P}_{k,q} $ at Sensor $ k $ is given by
		\begin{align}\label{hatP_to_phi}
			\hat{P}_{k,q}
			&=P_{\mathrm{t}}\left( 
			\frac{\kappa _{\mathrm{r}}\kappa _{\mathrm{t}}\beta _{\mathrm{r},k}\beta _{\mathrm{t},q}}{\left( \kappa _{\mathrm{r}}+1 \right) \left( \kappa _{\mathrm{t}}+1 \right)}{{\bm{\phi}} }_q^{\mathrm{H}}{{\bm \psi} }_{k,q}{{\bm \psi} }_{k,q}^{\mathrm{H}}{{\bm{\phi}} }_q\right.  \notag \\
			&\left. \quad+2\sqrt{\frac{\kappa _{\mathrm{d}}\kappa _{\mathrm{r}}\kappa _{\mathrm{t}}\beta _{\mathrm{d},k,q}\beta _{\mathrm{r},k}\beta _{\mathrm{t},q}}{\left( \kappa _{\mathrm{d}}+1 \right) \left( \kappa _{\mathrm{r}}+1 \right) \left( \kappa _{\mathrm{t}}+1 \right)}}\mathrm{Re}\left\{ {{\bm \psi} }_{k,q}^{\mathrm{H}}{{\bm{\phi}} }_q \right\}\right. \notag\\
			&\left. \quad +\beta _{\mathrm{d},k,q}+\frac{M\left( \kappa _{\mathrm{r}}+\kappa _{\mathrm{t}}+1 \right) \beta _{\mathrm{r},k}\beta _{\mathrm{t},q}}{\left( \kappa _{\mathrm{r}}+1 \right) \left( \kappa _{\mathrm{t}}+1 \right)}
			\right),
		\end{align}
		where 
		\begin{align}
			\psi_{k,m,q}&\triangleq \frac{2\pi}{\lambda} \left(  \left( d_{\mathrm{d},k,q}+d_{\mathrm{r},k}-d_{\mathrm{t},q} \right) \right. \notag \\
			&\left. \quad+d\left( \cos \omega _{\mathrm{r},k}-\cos \omega _{\mathrm{t},q} \right) \left( m-1 \right) \right),\label{varphi}
			\\
			{{\bm \psi} }_{k,q}&\triangleq \left[ e^{-j\psi _{k,1,q}},\dots ,e^{-j\psi _{k,M,q}} \right] ^{\mathrm{T}}.\label{Varphi}
		\end{align}

		\textit{Proof:} Please refer to Appendix \ref{appendix_hatP}. $\hfill\blacksquare$
	\end{theorem}
	The results in the above theorem will be used in the following sections for the system design.

	\section{Fly-Hover-Broadcast Protocol}\label{FLY-HOVER-BROADCAST PROTOCOL}
	The widely adopted FHB protocol is intuitive and easy to implement in practice, which also provides a performance lower bound for the general scenario. 
	In this section, we aim to minimize the total energy consumption of the UAV based on the FHB protocol.
	An efficient algorithm is developed to solve the formulated problem.
%	In this section, we formulate the energy consumption minimization problem based on the FHB protocol, for which an efficient algorithm is then developed.
	
	\subsection{Protocol and Energy Consumption Model}\label{section_sysmodel_FHB}
	Under the FHB protocol, it is assumed that the UAV radiates RF signals only when hovering.
	The trajectory consists of the initial and final locations $q_{\mathrm{I}}$ and $q_{\mathrm{F}}$, and $L-1$ hovering locations connected by $L$ straight path segments.
	We denote the $l$th hovering location by $q_{\mathrm{U},l}=q_{\mathrm{U},l}^{x}+jq_{\mathrm{U},l}^{y}$ for $l \in \mathcal{L} _{\mathrm{b}}$, where $\mathcal{L} _{\mathrm{b}}=\left\{ 1,\dots , L-1 \right\} $.
	On each flight, the UAV starts from $q_{\mathrm{I}}$, successively visits the hovering locations, and finally arrives at $q_{\mathrm{F}}$.
	Then, the total trajectory is denoted by $\mathbf{q}_{\mathrm{U}}=\left[ q_{\mathrm{U},0},\dots ,q_{\mathrm{U},L} \right] ^{\mathrm{T}}\in \mathbb{C} ^{\left( L+1 \right) \times 1}$ with %, where the constraints in \eqref{qI&qF} still hold.
	\begin{equation}\label{qI&qF}
		q_{\mathrm{U},0} = q_{\mathrm{I}}, \quad q_{\mathrm{U},L}=q_{\mathrm{F}}.
	\end{equation}
	It was derived in \cite{8663615} that each rotary-wing UAV has a \emph{maximum-range (MR) speed} ${v_{\mathrm{mr}}}$ that maximizes the total travel distance with any given onboard energy, which can be found from the plot of propulsion power $P_{\mathrm{p}}\left( v \right)$ versus UAV speed $ v $.
	With the MR speed $ {v_{\mathrm{mr}}} $, the flight time of the UAV along the $l$th path segment is given by $\frac{\left| q_{\mathrm{U},l}-q_{\mathrm{U},l-1} \right|}{v_{\mathrm{mr}}} $.
	By substituting $v={v_{\mathrm{mr}}}$ into \eqref{P_p}, we have the MR energy consumption $P_{\mathrm{p}}^{\mathrm{mr}}=P_{\mathrm{p}}\left( v_{\mathrm{mr}} \right)$.
	Similarly, the hovering energy consumption is derived as $P_{\mathrm{p}}^{\mathrm{hov}}=P_{\mathrm{p}}\left( 0 \right)=P_0+P_{\mathrm{i}}$.
	Denoting the hovering time of the UAV at the $l$th hovering location by $t_l$,
	we have the total energy consumption of the UAV as follows
	\begin{equation}\label{E_U_FHB}
		E_{\mathrm{U}}^{\mathrm{FHB}}\left( \mathbf{q}_{\mathrm{U}},\mathbf{t} \right) =P_{\mathrm{p}}^{\mathrm{mr}}\!\sum_{l=1}^L{\frac{\left| q_{\mathrm{U},l}-q_{\mathrm{U},l-1} \right|}{v_{\mathrm{mr}}}}+\sum_{l \in \mathcal{L} _{\mathrm{b}}}\!\!{\left( P_{\mathrm{p}}^{\mathrm{hov}}+P_{\mathrm{t}} \right) t_l},
	\end{equation}
	where $\mathbf{t}=\left[ t_1,\dots ,t_{L-1} \right] ^{\mathrm{T}}$.

	\subsection{Problem Formulation}\label{section_problem_formulation_FHB}
	When the UAV is at the $ l $th hovering location, we denote the phase shift of the $m$th reflecting element of the RIS by $\theta _{m,l}$.
	${{\bm{\phi}}}_l=\left[ e^{j\theta _{1,l}},...,e^{j\theta _{M,l}} \right] ^{\mathrm{T}}$ and $\mathbf{\Phi }_l={\mathrm{diag}}\left( {\bm{\phi}}_l \right) $ are also defined for $ l \in \mathcal{L} _{\mathrm{b}} $, which satisfies
	\begin{equation}\label{constraint_theta_FHB}
		\theta _{m,l}\in\left[ 0,2\pi \right) , \;m = 1, \dots,M, \;l \in \mathcal{L} _{\mathrm{b}}.
	\end{equation}
	Let $\beta _{\mathrm{t},l}$, $\beta _{\mathrm{d},k,l}$ and $ {{\bm \psi} }_{k,l} $ denote the values of $\beta _{\mathrm{t},q}$, $\beta _{\mathrm{d},k,q}$ and $ {{\bm \psi} }_{k,q} $ when the UAV is at the $l$th hovering location, respectively.
	From \eqref{hatP_to_phi}, the expectation of the received power at Sensor $k$ is given by
	\begin{align}\label{hatP_to_phi_FHB}
		\hat{P}_{k,l}
		&=P_{\mathrm{t}}\left( 
		\frac{\kappa _{\mathrm{r}}\kappa _{\mathrm{t}}\beta _{\mathrm{r},k}\beta _{\mathrm{t},l}}{\left( \kappa _{\mathrm{r}}+1 \right) \left( \kappa _{\mathrm{t}}+1 \right)}{{\bm{\phi}} }_l^{\mathrm{H}}{{\bm \psi} }_{k,l}{{\bm \psi} }_{k,l}^{\mathrm{H}}{{\bm{\phi}} }_l\right.  \notag \\
		&\left. \quad+2\sqrt{\frac{\kappa _{\mathrm{d}}\kappa _{\mathrm{r}}\kappa _{\mathrm{t}}\beta _{\mathrm{d},k,l}\beta _{\mathrm{r},k}\beta _{\mathrm{t},l}}{\left( \kappa _{\mathrm{d}}+1 \right) \left( \kappa _{\mathrm{r}}+1 \right) \left( \kappa _{\mathrm{t}}+1 \right)}}\mathrm{Re}\left\{ {{\bm \psi} }_{k,l}^{\mathrm{H}}{{\bm{\phi}} }_l \right\}\right. \notag\\
		&\left. \quad +\beta _{\mathrm{d},k,l}+\frac{M\left( \kappa _{\mathrm{r}}+\kappa _{\mathrm{t}}+1 \right) \beta _{\mathrm{r},k}\beta _{\mathrm{t},l}}{\left( \kappa _{\mathrm{r}}+1 \right) \left( \kappa _{\mathrm{t}}+1 \right)}
		\right).
	\end{align}

	In this section, we minimize the UAV energy consumption while
	providing the required energy for all sensors under the FHB protocol,
	via jointly optimizing the trajectory $\mathbf{q}_{\mathrm{U}}$,
	the hovering time $\mathbf{t}$
	and the reflection coefficient vectors $\left\{ {{\bm{\phi}} }_l \right\}$.
	The energy minimization problem can be formulated as follows
	\begin{subequations}\label{Prob1_FHB}
		\begin{alignat}{2}
			\min_{{\mathbf{q}_{\mathrm{U}},\mathbf{t},\left\{ {{\bm{\phi}} }_l \right\} }} \quad
			& E_{\mathrm{U}}^{\mathrm{FHB}}\left( \mathbf{q}_{\mathrm{U}},\mathbf{t} \right) \\
			\mbox{s.t.}\qquad
			&t_l\geqslant 0, \;l \in \mathcal{L} _{\mathrm{b}}, \label{constraint_t_FHB}\\
			&\eta \sum_{l \in \mathcal{L} _{\mathrm{b}}}{t_l\hat{P}_{k,l}}\geqslant E_{k}^{\mathrm{req}},\;k=1,\dots ,K,\label{constraint_E_Prob0_FHB} \\
			&\eqref{qI&qF}, \eqref{constraint_theta_FHB}.\notag
		\end{alignat}
	\end{subequations}
	The main notations used in Problem \eqref{Prob1_FHB} are summarized in Table \ref{Table_notation}.
	In Problem \eqref{Prob1_FHB}, the trajectory $\mathbf{q}_{\mathrm{U}}$, the hovering time $\mathbf{t}$
	and the reflection coefficient vectors $\left\{ {{\bm{\phi}} }_l \right\}$ are jointly optimized.
	By applying the MR speed ${v_{\mathrm{mr}}}$, the objective function is transformed to a convex function.
	However, constraints \eqref{constraint_theta_FHB} and  \eqref{constraint_E_Prob0_FHB} are still non-convex, and all variables are coupled in \eqref{constraint_E_Prob0_FHB}.
	In the following, we propose an efficient method for this complicated problem by decoupling it into two subproblems.

	\begin{table*}[tb]
		\begin{center}
				\caption{Main Notations and Terminologies under FHB Protocol and PD Protocol}
				\label{Table_notation}
				\begin{tabular}{c l l} 
					\hline\hline\noalign{\smallskip}
%					\textbf{Notation} & \multicolumn{2}{c}{\textbf{Definition}}\\		
%					& {\emph{FHB protocol}} & {\emph{PD protocol}} \\
					\textbf{Notation} & \textbf{Definition (FHB protocol)} & \textbf{Definition (PD protocol)} \\
					\hline\noalign{\smallskip}
					$L$ & Number of hovering locations, $\mathcal{L} _{\mathrm{b}}=\left\{1,\dots , L-1 \right\}$ & Number of path segments, $\mathcal{L} _{\mathrm{a}}=\left\{1,\dots , L \right\}$ \\
					$\mathbf{q}_{\mathrm{U}}$ & UAV's trajectory, $\mathbf{q}_{\mathrm{U}}=[ q_{\mathrm{U},0},\dots ,q_{\mathrm{U},L} ] ^{\mathrm{T}}$ & UAV's trajectory, $\mathbf{q}_{\mathrm{U}}=[ q_{\mathrm{U},0},\dots ,q_{\mathrm{U},L} ] ^{\mathrm{T}}$ \\
					$\mathbf{t}$ & UAV's hovering time, $\mathbf{t}=\left[ t_1,\dots ,t_{L-1} \right] ^{\mathrm{T}}$ & UAV's flight time, $\mathbf{t}=\left[ t_1,\dots ,t_{L} \right] ^{\mathrm{T}}$ \\
					$\{\bm{\phi}_l\}$ & RIS's reflection coefficients, $l\in\mathcal{L} _{\mathrm{b}}$ & RIS's reflection coefficients, $l\in\mathcal{L} _{\mathrm{a}}$ \\
					$\eta$ & Sensors' energy conversion efficiency & Sensors' energy conversion efficiency \\
					$ \hat{P}_{k,l}$ & Expected received power at Sensor $k$, $l\in\mathcal{L} _{\mathrm{b}}$ & Expected received power at Sensor $k$, $l\in\mathcal{L} _{\mathrm{a}}$\\
					$E_{k}^{\mathrm{req}}$ & Required energy for Sensor $k$ & Required energy for Sensor $k$ \\	
					\noalign{\smallskip}\hline
				\end{tabular}
		\end{center}
	\end{table*}

	\subsection{Optimizing the Trajectory $\mathbf{q}_{\mathrm{U}}$ and Hovering Time $\mathbf{t}$} \label{section_opt_qU_t_FHB}
	In this subsection, we propose an iterative algorithm based on the SCA method to jointly optimize $\mathbf{q}_{\mathrm{U}}$ and $\mathbf{t}$ given $\left\{{\bm{\phi}}_l\right\}$.
	From Problem \eqref{Prob1_FHB}, the subproblem of $\mathbf{q}_{\mathrm{U}}$ and $\mathbf{t}$ is formulated as follows
	\begin{subequations}\label{ProbU_FHB}
		\begin{alignat}{2}
			\min_{{\mathbf{q}_{\mathrm{U}},\mathbf{t}}} \quad
			& E_{\mathrm{U}}^{\mathrm{FHB}}\left( \mathbf{q}_{\mathrm{U}},\mathbf{t} \right) \\
			\mbox{s.t.}\quad
			&\eqref{qI&qF},\eqref{constraint_t_FHB},\eqref{constraint_E_Prob0_FHB}.\notag
		\end{alignat}
	\end{subequations}
	From \eqref{hatP_to_phi_FHB}, it can be found that $\hat{P}_{k,l}$ is extremely complex with respect to (w.r.t.) $\mathbf{q}_{\mathrm{U}}$,
	which makes the energy requirement constraint \eqref{constraint_E_Prob0_FHB} intractable.
	Fortunately, the change in the value of $ \mathbf{q}_{\mathrm{U}} $ per iteration is generally small.
	Denoting by $ {{\bm \psi}}_{k,l}^n $ the value of  $ {{\bm \psi}}_{k,l} $ at the $n$th iteration, the following approximation can be introduced with limited approximation error %(especially as the algorithm approaches convergence)
	\begin{equation}\label{hatP_approx}
		\hat{P}_{k,l}\approx P_{\mathrm{t}}\!\left( \left( U_{1,k,l}^{ n }+U_{3,k,l} \right) \beta _{\mathrm{t},l}+U_{2,k,l}^n\sqrt{\beta _{\mathrm{d},k,l}\beta _{\mathrm{t},l}}+\beta _{\mathrm{d},k,l} \right)\!,
	\end{equation}
	where $U_{1,k,l}^n$, $U_{2,k,l}^n$ and $U_{3,k,l}$ are respectively defined as
	\begin{align}
		&U_{1,k,l}^n=\frac{\kappa _{\mathrm{r}}\kappa _{\mathrm{t}}\beta _{\mathrm{r},k}\beta _{\mathrm{t},l}}{\left( \kappa _{\mathrm{r}}+1 \right) \left( \kappa _{\mathrm{t}}+1 \right)}{{\bm{\phi}}}_{l}^{\mathrm{H}}{{\bm \psi}}_{k,l}^n{{\bm \psi}}_{k,l}^{n ,\mathrm{H}}{{\bm{\phi}}}_l,
		\label{U1}\\
		&U_{2,k,l}^n=2\sqrt{\frac{\kappa _{\mathrm{d}}\kappa _{\mathrm{r}}\kappa _{\mathrm{t}}\beta _{\mathrm{d},k,l}\beta _{\mathrm{r},k}\beta _{\mathrm{t},l}}{\left( \kappa _{\mathrm{d}}+1 \right) \left( \kappa _{\mathrm{r}}+1 \right) \left( \kappa _{\mathrm{t}}+1 \right)}}\mathrm{Re}\left\{ {{\bm \psi}}_{k,l}^{n ,\mathrm{H}}{{\bm{\phi}}}_l \right\},
		\\
		&U_{3,k,l}=\frac{M\left( \kappa _{\mathrm{r}}+\kappa _{\mathrm{t}}+1 \right) \beta _{\mathrm{r},k}}{\left( \kappa _{\mathrm{r}}+1 \right) \left( \kappa _{\mathrm{t}}+1 \right)} \label{U3}.
	\end{align}
	Then, constraint \eqref{constraint_E_Prob0_FHB} is approximated as
	\begin{align}\label{constraint_E_approx_FHB}
		&\frac{\eta}{E_{k}^{\mathrm{req}}}\sum_{l \in \mathcal{L} _{\mathrm{b}}}P_{\mathrm{t}}t_l\big\{ \left( U_{1,k,l}^n+U_{3,k,l} \right) \beta _{\mathrm{t},l}+U_{2,k,l}^n\sqrt{\beta _{\mathrm{d},k,l}\beta _{\mathrm{t},l}} \notag \\
		& +\beta _{\mathrm{d},k,l} \big\}\geqslant 1, \;k=1,\dots ,K.
	\end{align}
	However, constraint \eqref{constraint_E_approx_FHB} is still non-convex.
	In the following, we utilize the SCA method to transform it to several convex constraints.
	Firstly, by using the fact that the first-order Taylor expansion is a lower bound of a convex function, we have the following inequalities
	\begin{align}\label{bar_beta_d}
		\beta _{\mathrm{d},k,l}&\geqslant \frac{\beta _0}{\left( \left| q_{\mathrm{U},l}^n-q_{\mathrm{S},k} \right|^2+H_{\mathrm{U}}^{2} \right) ^{\frac{\alpha _{\mathrm{d}}}{2}}} \notag \\
		&\quad-\frac{\alpha _{\mathrm{d}}\beta _0\left( \left| q_{\mathrm{U},l}-q_{\mathrm{S},k} \right|^2-\left| q_{\mathrm{U},l}^n-q_{\mathrm{S},k} \right|^2 \right)}{2\left( \left| q_{\mathrm{U},l}^n-q_{\mathrm{S},k} \right|^2+H_{\mathrm{U}}^{2} \right) ^{\frac{\alpha _{\mathrm{d}}}{2}+1}}
		\triangleq \bar{\beta}_{\mathrm{d},k,l},
	\end{align}
	\begin{align}\label{bar_beta_t}
		\!\!\beta _{\mathrm{t},l}&\geqslant \frac{\beta _0}{\left( \left| q_{\mathrm{U},l}^n-q_{\mathrm{R}} \right|^2+\left( H_{\mathrm{U}}-H_{\mathrm{R}} \right) ^2 \right) ^{\frac{\alpha _{\mathrm{t}}}{2}}}  \notag \\
		&\quad-\frac{\alpha _{\mathrm{t}}\beta _0\left( \left| q_{\mathrm{U},l}-q_{\mathrm{R}} \right|^2-\left| q_{\mathrm{U},l}^n-q_{\mathrm{R}} \right|^2 \right)}{2\left( \left| q_{\mathrm{U},l}^n-q_{\mathrm{R}} \right|^2+\left( H_{\mathrm{U}}-H_{\mathrm{R}} \right) ^2 \right) ^{\frac{\alpha _{\mathrm{t}}}{2}+1}}
		\triangleq \bar{\beta}_{\mathrm{t},l},
	\end{align}
	where $ q_{\mathrm{U},l}^n$ is the value of $ q_{\mathrm{U},l}$ at the $n$th iteration.
	Then, the problem to be solved at the $n$th SCA iteration can be formulated as follows
	\begin{subequations}\label{ProbU_SCA1_FHB}
		\begin{alignat}{2}
			\min_{\mathbf{q}_{\mathrm{U}},\mathbf{t},\{ y_{\mathrm{t},l} \} ,
				\atop
				\{ y_{\mathrm{d},k,l} \} } \quad
			& E_{\mathrm{U}}^{\mathrm{FHB}}\left( \mathbf{q}_{\mathrm{U}},\mathbf{t} \right) \\
			\mbox{s.t.}\qquad\;
			&\frac{\eta}{E_{k}^{\mathrm{req}}}\sum_{l \in \mathcal{L} _{\mathrm{b}}}P_{\mathrm{t}}t_l\big( \left( U_{1,k,l}^n+U_{3,k,l} \right) y_{\mathrm{t},l} \notag \\
			& +U_{2,k,l}^n\sqrt{y_{\mathrm{t},l}y_{\mathrm{d},k,l}}+y_{\mathrm{d},k,l} \big)\ge 1, \;k=1,\dots ,K,\label{constraint_E_ProbU_SCA1_FHB}
			\\
			&\bar{\beta}_{\mathrm{t},l}\geqslant y_{\mathrm{t},l}\geqslant 0,  \;l \in \mathcal{L} _{\mathrm{b}},\label{constraint_yt_ProbU_SCA_FHB}
			\\
			&\bar{\beta}_{\mathrm{d},k,l}\geqslant y_{\mathrm{d},k,l}\geqslant 0, \;k=1,\dots ,K,\label{constraint_yd_ProbU_SCA_FHB}
			\\
			&\eqref{qI&qF},\eqref{constraint_t_FHB}.\notag
		\end{alignat}
	\end{subequations}
	Note that \eqref{constraint_E_ProbU_SCA1_FHB} is still non-convex,
	we introduce two sets of slack variables $\left\{ y_{\mathrm{a},k,l} \right\}$ and $\left\{ e_{k,l} \right\}$ to transform it as the following constraints
	\begin{align}\label{constraint_E_ProbU_SCA2_FHB}
		&\left( U_{1,k,l}^n+U_{3,k,l} \right) y_{\mathrm{t},l}+U_{2,k,l}^ny_{\mathrm{a},k,l}+y_{\mathrm{d},k,l}\geqslant \frac{E_{k}^{\mathrm{req}}e_{k,l}^{2}}{\eta P_{\mathrm{t}}t_l}, \notag \\
		&\;{l \in \mathcal{L} _{\mathrm{b}}} ,\;k=1,\dots ,K,
	\end{align}
	\begin{equation}\label{constraint_ya_ProbU_SCA_FHB}
		y_{\mathrm{t},l}\geqslant \frac{y_{a,k,l}^{2}}{y_{\mathrm{d},k,l}}, \;{l \in \mathcal{L} _{\mathrm{b}}},\;k=1,\dots ,K,
	\end{equation}
	\begin{equation}\label{constraint_e_ProbU_FHB}
		\sum_{l \in \mathcal{L} _{\mathrm{b}}}{e_{k,l}^{2}}\geqslant 1,\;k=1,\dots ,K,
	\end{equation}
	and utilize the first-order Taylor expansion to transform \eqref{constraint_e_ProbU_FHB} as
	\begin{equation}\label{constraint_e_ProbU_SCA_FHB}
		\sum_{l \in \mathcal{L} _{\mathrm{b}}}{\left( 2e_{k,l}^ne_{k,l}-e_{k,l}^{2,n} \right)}\geqslant 1,\;k=1,\dots ,K,
	\end{equation}
	where $e_{k,l}^n$ is the value of $e_{k,l}$ at the $n$th iteration.
	Finally, Problem \eqref{ProbU_SCA1_FHB} is reformulated as
	\begin{subequations}\label{ProbU_SCA2_FHB}
		\begin{alignat}{2}
			\min_{\mathbf{q}_{\mathrm{U}},\mathbf{t},\{ y_{\mathrm{t},l} \},\{ y_{\mathrm{d},k,l} \},
				\atop
				\{ y_{\mathrm{a},k,l} \} ,\{ e_{k,l} \} } \quad
			& E_{\mathrm{U}}^{\mathrm{FHB}}\left( \mathbf{q}_{\mathrm{U}},\mathbf{t} \right) \\
			\mbox{s.t.}\qquad\qquad\!
			&\eqref{qI&qF},\eqref{constraint_t_FHB},\eqref{constraint_yt_ProbU_SCA_FHB},
			\eqref{constraint_yd_ProbU_SCA_FHB},\eqref{constraint_E_ProbU_SCA2_FHB},\eqref{constraint_ya_ProbU_SCA_FHB},\eqref{constraint_e_ProbU_SCA_FHB}.\notag
		\end{alignat}
	\end{subequations}
	Problem \eqref{ProbU_SCA2_FHB} is a convex problem that can be efficiently solved by the 	existing optimization tools, such as CVX.
	
	\subsection{Optimizing the Reflection Coefficient Vectors $\left\{{\bm{\phi}}_l\right\}$}\label{section_opt_phi_FHB}
	In this subsection, we optimize $\left\{{\bm{\phi}}_l\right\}$ given $\mathbf{q}_{\mathrm{U}}$ and $\mathbf{t}$.
	Note that the objective function of \eqref{Prob1_FHB} is independent of $\left\{{\bm{\phi}}_l\right\}$.
	Hence, the subproblem of $\left\{{\bm{\phi}}_l\right\}$ is a feasibility-check problem.
	To improve convergence performance, a common approach is to strengthen the optimization objective to maximize the total oversupplied energy of all sensors \cite{9180053,8811733}.
	However, due to the energy requirement constraint \eqref{constraint_E_Prob0_FHB}, the transformed subproblem is not guaranteed to be feasible, which may lead to early convergence of the algorithm.
	To address this, we set a more challenging optimization objective to maximize the minimum charged energy of all sensors instead.
	The corresponding subproblem is formulated as follows
	\begin{subequations}\label{ProbR_FHB}
		\begin{alignat}{2}
			\max_{{\left\{ {{\bm{\phi}} }_l \right\},\epsilon }} \quad
			& \epsilon \\
			\mbox{s.t.}\quad \;
			&\frac{\eta}{E_{k}^{\mathrm{req}}} \sum_{l \in \mathcal{L} _{\mathrm{b}}}{t_l \hat P_{k,l}}\geqslant \epsilon,\;k=1,\dots ,K,\\
			&\eqref{constraint_theta_FHB},\notag
		\end{alignat}
	\end{subequations}
	where $\epsilon$ is an auxiliary variable.
	To guarantee the feasibility, constraint \eqref{constraint_E_Prob0_FHB} in Problem \eqref{ProbR_FHB} is relaxed by removing the constraint $\epsilon\geqslant 1$.
	In this paper, we propose to obtain the solution of Problem \eqref{Prob1_FHB} by solving problems \eqref{ProbU_SCA2_FHB} and  \eqref{ProbR_FHB} alternately.
	Since the former involves the energy requirement constraint \eqref{constraint_E_ProbU_SCA2_FHB}, the relaxation technique will not change the feasibility of the converged solution in Problem \eqref{Prob1_FHB}.
	%the converged solution of our algorithm is ensured to satisfy constraint \eqref{constraint_E_Prob0_FHB}.
	Note that constraint \eqref{constraint_theta_FHB} imposes $ M\left( L-1\right) $ unit-modulus constraints on $ \left\{ {{\bm{\phi}} }_l \right\} $.
	For schemes with large $ L $, Gaussian randomization in the SDR-based method incurs significant performance loss when solving Problem \eqref{ProbR_FHB}.\footnote{Under general protocols, such as the PD protocol applied in Section \ref{PATH DISCRETIZATION PROTOCOL}, the value of $ L $ is much larger and the performance loss is more severe. }		
	In the following, we propose a low-complexity algorithm to solve Problem \eqref{ProbR_FHB} based on the widely used MM method \cite{zhou2019intelligent,9318531,hunter2004tutorial,7547360}.
	
	To begin with, we reformulate Problem \eqref{ProbR_FHB} as a more tractable form.
	By defining 
	\begin{align}
		\mathbf{A}_{k,l}&\triangleq P_{\mathrm{t}}\frac{\kappa _{\mathrm{r}}\kappa _{\mathrm{t}}\beta _{\mathrm{r},k}\beta _{\mathrm{t},l}}{\left( \kappa _{\mathrm{r}}+1 \right) \left( \kappa _{\mathrm{t}}+1 \right)}{{\bm \psi} }_{k,l}{{\bm \psi} }_{k,l}^{\mathrm{H}},
		\\
		\mathbf{a}_{k,l}&\triangleq P_{\mathrm{t}}\sqrt{\frac{\kappa _{\mathrm{d}}\kappa _{\mathrm{r}}\kappa _{\mathrm{t}}\beta _{\mathrm{d},k,l}\beta _{\mathrm{r},k}\beta _{\mathrm{t},l}}{\left( \kappa _{\mathrm{d}}+1 \right) \left( \kappa _{\mathrm{r}}+1 \right) \left( \kappa _{\mathrm{t}}+1 \right)}}{{\bm \psi} }_{k,l},
		\\
		cons\theta _{k,l}&\triangleq P_{\mathrm{t}}\left( \beta _{\mathrm{d},k,l}+\frac{M\left( \kappa _{\mathrm{r}}+\kappa _{\mathrm{t}}+1 \right) \beta _{\mathrm{r},k}\beta _{\mathrm{t},l}}{\left( \kappa _{\mathrm{r}}+1 \right) \left( \kappa _{\mathrm{t}}+1 \right)} \right),
	\end{align}
	we can transform \eqref{hatP_to_phi} as follows
	\begin{align}
		\hat{P}_{k,l}
		= {{\bm{\phi}} }_{l}^{\mathrm{H}}\mathbf{A}_{k,l}{\bm{\phi}}_l+2\mathrm{Re}\left\{ \mathbf{a}_{k,l}^{\mathrm{H}}{\bm{\phi}}_l \right\} +cons\theta _{k,l}.
	\end{align}
	Then, we define $h_k\left( {\bm{\phi}} \right) \triangleq \frac{\eta}{E_{k}^{\mathrm{req}}}\sum_{l \in \mathcal{L} _{\mathrm{b}}}{t_l\hat{P}_{k,l}}$ and transform it into the following quadratic form
	\begin{align}\label{hphi}
		h_k\left( {\bm{\phi}} \right)&=\sum_{l \in \mathcal{L} _{\mathrm{b}}}{\frac{\eta t_l}{E_{k}^{\mathrm{req}}}{\bm{\phi}}_{l}^{\mathrm{H}}\mathbf{A}_{k,l}{\bm{\phi}}_l}+2\sum_{l \in \mathcal{L} _{\mathrm{b}}}{\frac{\eta t_l}{E_{k}^{\mathrm{req}}}\mathrm{Re}\left\{ \mathbf{a}_{k,l}^{\mathrm{H}}{\bm{\phi}}_l \right\}} \notag \\
		&\quad+\sum_{l \in \mathcal{L} _{\mathrm{b}}}{\frac{\eta t_l}{E_{k}^{\mathrm{req}}}cons\theta _{k,l}} \notag
		\\
		&={\bm{\phi}}^{\mathrm{H}}\mathbf{B}_k{\bm{\phi}}+2\mathrm{Re}\left\{ \mathbf{b}_{k}^{\mathrm{H}}{\bm{\phi}} \right\} +cons\phi _k,
	\end{align}
	where
	\begin{align}
		{\bm{\phi}}&\triangleq \left[ {\bm{\phi}}_{1}^{\mathrm{T}},\dots ,{\bm{\phi}}_{L-1}^{\mathrm{T}} \right] ^{\mathrm{T}},
		\\
		\mathbf{B}_k&\triangleq \left[ \begin{matrix}
			\frac{\eta t_1}{E_{k}^{\mathrm{req}}}\mathbf{A}_{k,1}&		&		\\
			&		\ddots&		\\
			&		&		\frac{\eta t_{L-1}}{E_{k}^{\mathrm{req}}}\mathbf{A}_{k,L-1}\\
		\end{matrix} \right],\label{B_k}
		\\
		\mathbf{b}_k & \triangleq \left[ \frac{\eta t_1}{E_{k}^{\mathrm{req}}}\mathbf{a}_{k,1}^{\mathrm{T}},\dots ,\frac{\eta t_{L-1}}{E_{k}^{\mathrm{req}}}\mathbf{a}_{k,L-1}^{\mathrm{T}} \right] ^{\mathrm{T}},\label{b_k}
		\\
		cons\phi _k&\triangleq \sum_{l \in \mathcal{L} _{\mathrm{b}}}{\frac{\eta t_l}{E_{k}^{\mathrm{req}}}cons\theta _{k,l}}.
	\end{align}
	Finally, Problem \eqref{ProbR_FHB} is reformulated as follows
	\begin{subequations}\label{ProbR2_FHB}
		\begin{alignat}{2}
			\max_{{\bm{\phi}}} \quad
			& {\min_k}\left\{ h_k\left( {\bm{\phi}} \right) \right\} \\
			\mbox{s.t.}\quad
			&\eqref{constraint_theta_FHB}\notag
		\end{alignat}
	\end{subequations}

	Note that the objective function of Problem \eqref{ProbR2_FHB} is non-differentiable.
	To address this, we replace it with the following smooth and convex lower bound \cite{Xu2001Smoothing}
	\begin{equation}\label{smooth_approx}
		\!{\min_k}\left\{ h_k\left( {\bm{\phi}} \right) \right\} \!\approx\! f\left( {\bm{\phi}} \right) \!\triangleq\! -\frac{1}{\mu} \log\! \left( \sum_{k=1}^K{\exp \left( -\mu h_k\left( {\bm{\phi}} \right) \right)}\! \right) \!,\!
	\end{equation}
	where $ \mu > 0 $ is a smoothing parameter.
	In particular, the following inequalities hold
	\begin{equation}\label{smooth_approx_gap}
		f\left( {\bm{\phi}} \right) \leqslant {\min_k}\left\{ h_k\left( {\bm{\phi}} \right) \right\} \leqslant f\left( {\bm{\phi}} \right) +\frac{1}{\mu}\log \left( K \right) .
	\end{equation}
	By replacing the objective function of Problem \eqref{ProbR2_FHB} with $ f\left( {\bm{\phi}} \right) $, a convex problem can be obtained.
	To utilize the MM method, the following theorem provides a tractable miniorizing function of $ f\left( {\bm{\phi}} \right) $ to formulate the surrogate problem at each iteration.
	\begin{theorem}\label{theorem_minorizing}
		Denote by $ {\bm{\phi}}^r $ the solution at the $ r $th iteration.
		For any feasible ${\bm{\phi}}$, $ f\left( {\bm{\phi}} \right) $ is minorized with $ \tilde{f}\left( {\bm{\phi}}\mid {\bm{\phi}}^r \right) $ as follows 
		\begin{equation}\label{tildef_FHB}
			\tilde{f}\left( {\bm{\phi}}\mid {\bm{\phi}}^r \right) = 2\mathrm{Re}\left\{ \mathbf{u}^{\mathrm{H}}{\bm{\phi}} \right\} +cons\phi _{\mathrm{MM}}.
		\end{equation}
		where
		\begin{align}
			&g_k\left( {\bm{\phi}}^r \right) \triangleq \frac{\exp \left( -\mu h_k\left( {\bm{\phi}}^r \right) \right)}{\sum_{k=1}^K{\exp \left( -\mu h_k\left( {\bm{\phi}}^r \right) \right)}},\label{g_k_phir}
			\\
			&\mathbf{c}=\sum_{k=1}^K{g_k\left( {\bm{\phi}}^r \right) \left( \mathbf{B}_{k}^{\mathrm{H}}{\bm{\phi}}^r+\mathbf{b}_k \right)},
			\\
			&\alpha=-2\mu {\max_k} \Big\{M{\max_l}\left\{ \lambda _{\max}\left( \left( \frac{\eta t_l}{E_{k}^{\mathrm{req}}} \right) ^2\mathbf{A}_{k,l}\mathbf{A}_{k,l}^{\mathrm{H}} \right) \right\} \notag \\
			&\quad+\mathbf{b}_{k}^{\mathrm{H}}\mathbf{b}_k+2\left\| \mathbf{B}_k\mathbf{b}_k \right\| _1 \Big\},\label{alpha}
			\\
			&\mathbf{u}\triangleq \mathbf{c}-\alpha {\bm{\phi}}^r,\label{u}
			\\
			&cons\phi _{\mathrm{MM}}\triangleq f\left( {\bm{\phi}}^r \right) -2\mathrm{Re}\left\{ \mathbf{c}^{\mathrm{H}}{\bm{\phi}}^r \right\} +2\alpha M.
		\end{align}
		
		\textit{Proof:} Please refer to Appendix \ref{appendix_minorizing}. $\hfill\blacksquare$
	\end{theorem}
	
	By replacing the objective function of Problem \eqref{ProbR2_FHB} with \eqref{tildef_FHB}, we obtain the surrogate problem at the $r$th MM iteration
	\begin{subequations}\label{ProbR_Surrogate_FHB}
		\begin{alignat}{2}
			\max_{{\bm{\phi}}} \quad
			& 2\mathrm{Re}\left\{ \mathbf{u}^{\mathrm{H}}{{\bm{\phi}} } \right\} +cons\phi _{\mathrm{MM}}\\
			\mbox{s.t.}\quad
			&\eqref{constraint_theta_FHB}.\notag
		\end{alignat}
	\end{subequations}
	It can be readily verified that the optimal solution of Problem \eqref{ProbR_Surrogate_FHB} is given by
	\begin{equation}\label{phi^r+1}
		{{\bm{\phi}} }^{r+1}=\exp \left( j\angle \mathbf{u} \right),
	\end{equation}
	where $\exp \left( \cdot \right) $ and $\angle \left( \cdot \right) $ are element-wise operations.
	
	\begin{algorithm}[t] %算法开始
		\caption{MM algorithm for solving Problem \eqref{ProbR_FHB}} %算法的题目
		\label{Alg_MM} %算法的标签
		\begin{algorithmic}[1] %此处的[1]控制一下算法中的每句前面都有标号
			\Require  Initial the number of iterations as  $r=1$. Set feasible ${{\bm{\phi}}}^{1}$, smoothing parameter $\mu$, maximum number of iterations $r_{\max}$ and error tolerance $\varepsilon _{\mathrm{e}}$.
			\Repeat
			\State Calculate ${\hat{\bm{\phi}}}_1=\mathfrak{M} \left( {{\bm{\phi}}}^r \right) $ and ${\hat{\bm{\phi}}}_2=\mathfrak{M} \left( {\hat{\bm{\phi}}}_1 \right) $;\label{cal_phi1phi2}
			\State Calculate $\mathbf{v}_1={\hat{\bm{\phi}}}_1-{{\bm{\phi}}}^r$ and $\mathbf{v}_2={\hat{\bm{\phi}}}_2-{\hat{\bm{\phi}}}_1-\mathbf{v}_1$;
			\State Calculate step factor $\sigma =-\frac{\left\| \mathbf{v}_1 \right\| _2}{\left\| \mathbf{v}_2 \right\| _2}$;
			\State Calculate ${{\bm{\phi}}}^{r+1}=\exp \left( j\angle \left( {{\bm{\phi}}}^r-2\sigma \mathbf{v}_1+\sigma ^2\mathbf{v}_2 \right) \right) $;
			\State \textls[-5]{If $f\left( {{\bm{\phi}}}^{r+1} \right) <f\left( {{\bm{\phi}}}^r \right) $, set $\sigma \gets \frac{\left( \sigma -1 \right)}{2}$ and go to step \ref{cal_phi1phi2};}
			\State Set $r\gets r+1$;
			\Until ${\left| f\left( {{\bm{\phi}}}^{r+1} \right) -f\left( {{\bm{\phi}}}^r \right) \right|}<\varepsilon _{\mathrm{e}}{f\left( {{\bm{\phi}}}^r \right)}$ or $r\geqslant r_{\max}$;	
			\State If ${\min_k}\left\{ h_k\left( {\bm{\phi}}^{ r+1 } \right) \right\} < {\min_k}\left\{ h_k\left( {{\bm{\phi}}}^{ r } \right) \right\}$, set ${\bm{\phi}}^{ r+1 } \gets {{\bm{\phi}}}^1$.\label{abandon_solution}
		\end{algorithmic}
	\end{algorithm}
	
	Based on the above discussions,
	a modified MM algorithm is proposed to solve Problem \eqref{ProbR_FHB},
	the details of which are provided in Algorithm \ref{Alg_MM}, 
	where SQUAREM \cite{varadhan2008simple} theory is introduced to accelerate the convergence.
	Specifically, ${{\bm{\phi}}}^{r+1} =\mathfrak{M} \left( {{\bm{\phi}}}^r \right) $ in Step \ref{cal_phi1phi2} represents the nonlinear fixed-point iteration map given in \eqref{phi^r+1}.
	Thanks to the guaranteed feasibility of Problem \eqref{ProbR_FHB}, $\bm{\phi}$ can be efficiently optimized by Algorithm \ref{Alg_MM} with arbitrary $\mathbf{q}_{\mathrm{U}}$ and $\mathbf{t}$.
	%	As it will be shown in the next subsection, Algorithm \ref{Alg_MM} constitutes the inner iteration of the entire algorithm for solving Problem \eqref{Prob1_FHB}.
	Furthermore, the works in \cite{9090356}, \cite{zhou2019intelligent} and \cite{7547360} have proved and verified the Karush-Kuhn-Tucker (KKT) optimality of the converged solution of MM algorithm,
	which indicates that Algorithm \ref{Alg_MM} can obtain a KKT optimal solution of the relaxed Problem \eqref{ProbR2_FHB} with objective function $ f\left( {\bm{\phi}} \right) $.
	Moreover, when $ \mu $ is sufficiently large, the approximation in \eqref{smooth_approx} is tight, and the converged solution is approximately KKT optimal for Problem \eqref{ProbR_FHB}.

		\begin{algorithm}[t]
		\caption{SCA-MM algorithm under the FHB protocol}
		\label{Alg_SCA-MM_FHB}
		\begin{algorithmic}[1] %此处的[1]控制一下算法中的每句前面都有标号
			\Require  Initial the number of iterations as $n=1$. Set feasible ${{\bm{\phi}}}^1$, $\mathbf{q}_{\mathrm{U}}^1$ and $\mathbf{t}^1$, the maximum number of iterations $n_{\max}$, and smoothing-related factors $\mu,\mu _{\max}$ and $\iota$.
			\Repeat
			\State Given $\mathbf{q}_{\mathrm{U}}^n$, calculate $U_{1,k,l}^n$, $U_{2,k,l}^n$ and $U_{3,k,l}$ in \eqref{U1}-\eqref{U3};
			\State Set $\{ e_{k,l} ^n\}$ to hold the equalities in \eqref{constraint_E_ProbU_SCA2_FHB};
			\State Given ${{\bm{\phi}}}^n$, calculate the optimal $\mathbf{q}_{\mathrm{U}}^{n+1}$ and $\mathbf{t}^{ n+1 }$ by solving Problem \eqref{ProbU_SCA2_FHB};\label{update_qU_T_SCA-MM_FHB}
			%			\State Given $\mathbf{q}_{\mathrm{U}}^{n+1}$, update $ {{\bm \psi}}_{k,l} $ in \eqref{Varphi};
			\State Given $\mathbf{q}_{\mathrm{U}}^{n+1}$ and $\mathbf{t}^{ n+1 }$, calculate the optimal ${{\bm{\phi}}}^{ n+1 }$ with Algorithm \ref{Alg_MM};\label{update_phi_SCA-MM_FHB}
			\State Set $\mu \gets \max \left\{ \mu ^{\iota},\mu _{\max} \right\} $ and $n\gets n+1$;\label{adjust_mu_FHB}
			\Until $n\geqslant n_{\max}$.
		\end{algorithmic}
	\end{algorithm}
	
	\subsection{Algorithm Development}
	Based on the above discussions, we propose an effective and efficient algorithm for solving Problem \eqref{Prob1_FHB} in Algorithm \ref{Alg_SCA-MM_FHB}, named \textit{SCA-MM}.
%	where the SCA framework is conceived as the outer iteration.
	In Step \ref{update_phi_SCA-MM_FHB} of Algorithm \ref{Alg_SCA-MM_FHB}, Algorithm \ref{Alg_MM} is utilized to calculate the optimal ${{\bm{\phi}}}^{ n+1 }$,
	where ${{\bm{\phi}}}^{ n }$ in Algorithm \ref{Alg_SCA-MM_FHB} is chosen as the initial feasible ${{\bm{\phi}}}^{1}$ of Algorithm \ref{Alg_MM}.
	In the later stage of Algorithm \ref{Alg_SCA-MM_FHB}, a large $ \mu $ is required to improve the approximation accuracy  of $ f\left( {\bm{\phi}} \right) $ in \eqref{smooth_approx_gap}.
	On the contrary, a large $ \mu $ may result in premature convergence to a local stationary point.
	To take into account both the solution's optimality and the convergence speed, an adjustment factor $ \iota $ is introduced in Step \ref{adjust_mu_FHB}, which gradually increases $ \mu $ during iterations.
	Recall that the approximation \eqref{hatP_approx} is introduced in constraint \eqref{constraint_E_ProbU_SCA2_FHB}.
	Hence, the optimal solution to Problem \eqref{ProbU_SCA2_FHB} is not guaranteed to satisfy constraint \eqref{constraint_E_Prob0_FHB}.
	Although this issue will be addressed when Problem \eqref{ProbU_SCA2_FHB} is solved again in a subsequent iteration,
	Algorithm \ref{Alg_SCA-MM_FHB} may not generate monotonically decreasing objective function value of Problem \eqref{Prob1_FHB}.
	However, the simulations of convergence in Section \ref{SIMULATION RESULTS} show that the fluctuations are not significant.

	In the following, we analyze the computational complexity of Algorithm \ref{Alg_SCA-MM_FHB}, which mainly lies in solving Problem \eqref{ProbU_SCA2_FHB} in Step \ref{update_qU_T_SCA-MM_FHB} and calculating ${{\bm{\phi}}}^{ n+1 }$ in Step \ref{update_phi_SCA-MM_FHB}.
	Firstly, since Problem \eqref{ProbU_SCA2_FHB} involves only linear matrix inequality (LMI) and second-order cone (SOC) constraints, it can be solved by a standard interior point method \cite{Aharon2001Lectures}, whose computational complexity is given by
	\begin{equation*}
		\mathcal{O} \!\!\left(\! \sqrt{\underbrace{\sum_{j=1}^J{b_j}}_{\text{LMI}}+{\underbrace{2I}_{\text{SOC}}}} \left( {\underbrace{n^2\sum_{j=1}^J{b_{j}^{2}}+n\sum_{j=1}^J{b_{j}^{3}}}_{\text{LMI}}}+{\underbrace{n\sum_{i=1}^I{a_{i}^{2}}}_{\text{SOC}}}+n^3 \right)\!\! \right) \!,
	\end{equation*}
	where $n$ represents the number of variables, $J$ and $I$  denote the number of LMI and SOC constraints with sizes $\left\lbrace b_{j}\right\rbrace $ and $\left\lbrace a_{i}\right\rbrace $, respectively.
	Based on the above general expression, the complexity order of solving Problem \eqref{ProbU_SCA2_FHB} can be derived as $\mathcal{O}^{\mathrm{Alg. 2}}_{\mathrm{S}4}=\mathcal{O} ( \sqrt{\left( 5K+3 \right) \left( L-1 \right) +2K}( n^2K( L-1 ) ^2+n K( L-1 ) ^3 +n^3 ) )$, where $n= K\left( L-1 \right)$.
	Secondly, in each iteration of Algorithm \ref{Alg_MM}, the computational complexity depends on the calculation of $\alpha$ and the values of $h_k\left( {\bm{\phi}}^r \right)$ for $k=1,\dots ,K$.
	The corresponding complexity order are given by $\mathcal{O} ( K( ( L-1 ) ^2M^2+M^3 ) )$ and $\mathcal{O} ( K( ( L-1 ) ^2M^2+( L-1 ) M ) )$, respectively.
	Denote the number of iterations required for Algorithm \ref{Alg_MM} to converge by $r_{\mathrm{MM}}$.
	Then, the complexity of calculating ${{\bm{\phi}}}^{ n+1 }$ is of order $\mathcal{O}^{\mathrm{Alg. 2}}_{\mathrm{S}5}=\mathcal{O} ( r_{\mathrm{MM}}K( ( L-1 ) ^2M^2+M^3 ) ) $.
	Finally, the overall complexity of Algorithm \ref{Alg_SCA-MM_FHB} is given by $\mathcal{O}^{\mathrm{Alg. 2}}_{\mathrm{S}4}+\mathcal{O}^{\mathrm{Alg. 2}}_{\mathrm{S}5}$.

	\section{Path Discretization Protocol}\label{PATH DISCRETIZATION PROTOCOL}
	In order to maximize the harvested energy, the RF signals should be radiated during the UAV's flight, which is not considered in the FHB protocol for simplicity.
	In this section, the general scenario is studied under the PD protocol.
	Specifically, we solve the corresponding problem by extending the proposed method in the previous section.
	
	\subsection{Protocol and Energy Consumption Model}\label{section_sysmodel_PD}
	A common method to deal with the continuous flying trajectory of UAV is \emph{time discretization}, where the predetermined duration of entire flight is divided into several time slots.
	However, this approach is not applicable to our work due to the UAV energy consumption, which is related to the flight time.
	To address this issue, we adopt the PD method.
	%	Specifically, denoting $q_{\mathrm{I}}=q_{\mathrm{I}}^{x}+jq_{\mathrm{I}}^{y}$ and $q_{\mathrm{F}}=q_{\mathrm{F}}^{x}+jq_{\mathrm{F}}^{y}$ as the initial and final locations of the UAV, respectively,
	%	the UAV trajectory is discretized into $L$ path segments that are unequal in length. 
	%	When the $l$th path segment is sufficiently short, 
	%	it can be approximated as the line between its starting point $q_{\mathrm{U},l-1}$ and ending point $q_{\mathrm{U},l}$, where $q_{\mathrm{U},l}=q_{\mathrm{U},l}^{x}+jq_{\mathrm{U},l}^{y}$.
	%	Then, the UAV trajectory is represented by $\mathbf{q}_{\mathrm{U}}=\left[ q_{\mathrm{U},0},\dots ,q_{\mathrm{U},L} \right] ^{\mathrm{T}}\in \mathbb{C} ^{\left( L+1 \right) \times 1}$,
	%	where
	Specifically, the UAV trajectory is discretized into $L$ path segments that are unequal in length. 
	When the $l$th path segment is sufficiently short, 
	it can be approximated as the line between its starting point $q_{\mathrm{U},l-1}$ and ending point $q_{\mathrm{U},l}$, where $q_{\mathrm{U},l}=q_{\mathrm{U},l}^{x}+jq_{\mathrm{U},l}^{y}$.
	Then, the total trajectory is denoted by $\mathbf{q}_{\mathrm{U}}=\left[ q_{\mathrm{U},0},\dots ,q_{\mathrm{U},L} \right] ^{\mathrm{T}}\in \mathbb{C} ^{\left( L+1 \right) \times 1}$, where the constraints in \eqref{qI&qF} still hold.

	The following constraints are introduced on the length of path segments $\left\{\delta _l \right\}$
	\begin{equation}
		\delta_l \triangleq \left| q_{\mathrm{U},l}-q_{\mathrm{U},l-1} \right|\leqslant \varDelta _{\max}, \;l \in \mathcal{L} _{\mathrm{a}},
	\end{equation}
	where $\varDelta _{\max}$ is the maximum length of each path segment and $\mathcal{L} _{\mathrm{a}}=\left\{ 1,\dots ,L \right\} $.
	It is assumed that the UAV flies at a constant speed at each path segment.
	Due to the high mobility of rotary-wing UAV, the acceleration and deceleration process is ignored.
	In addition, the value of $\varDelta _{\max}$ should be chosen appropriately so that the distance from the UAV to each sensor and the RIS is approximately fixed at each path segment.
	Thus, the coordinates of any point on the $l$th path segment are regarded as $q_{\mathrm{U},l}$.
	
	Define $ t_l $ as the flight time of the UAV along the $l$th path segment.
	From \eqref{P_p}, the UAV's propulsion power at the $l$th path segment can be expressed as follows
	\begin{align}\label{P_p2}
		P_{\mathrm{p},l}\left( t_l \right)&=P_0\left( 1+\frac{3\delta _{l}^{2}}{U_{\mathrm{tip}}^{2}t_{l}^{2}} \right) +P_{\mathrm{i}}\left( \sqrt{1+\frac{\delta _{l}^{4}}{4v_{0}^{4}t_{l}^{4}}}-\frac{\delta _{l}^{2}}{2v_{0}^{2}t_{l}^{2}} \right) ^{\frac{1}{2}} \notag \\
		&\quad+\frac{1}{2}d_0\rho sA\frac{\delta _{l}^{3}}{t_{l}^{3}}.
	\end{align}
	Then, the total UAV energy consumption of one flight is given by
	\begin{equation}\label{E_U}
		E_{\mathrm{U}}\left( \mathbf{q}_{\mathrm{U}},\mathbf{t} \right) = \sum_{l \in \mathcal{L} _{\mathrm{a}}}{t_l\left( P_{\mathrm{t}}+P_{\mathrm{p},l}\left( t_l \right) \right)},
	\end{equation}
	where $\mathbf{t}=\left[ t_1,\dots ,t_{L} \right] ^{\mathrm{T}}$.

	\subsection{Problem Formulation}
	Similar to the FHB protocol, when the UAV is at the $ l $th path segment, we denote the phase shift of the $m$th reflecting element of the RIS by $\theta _{m,l}$,
	and introduce the definitions ${{\bm{\phi}}}_l=\left[ e^{j\theta _{1,l}},...,e^{j\theta _{M,l}} \right] ^{\mathrm{T}}$ and $\mathbf{\Phi }_l={\mathrm{diag}}\left( {\bm{\phi}}_l \right) $ for $ l \in \mathcal{L} _{\mathrm{a}} $, which satisfies
	\begin{equation}\label{constraint_theta_PD}
		\theta _{m,l}\in\left[ 0,2\pi \right) , \;m = 1, \dots,M, \;l \in \mathcal{L} _{\mathrm{a}}.
	\end{equation}
	Additionally, we denote the values of $\beta _{\mathrm{t},q}$, $\beta _{\mathrm{d},k,q}$ and $ {{\bm \psi} }_{k,q} $ when the UAV is at the $l$th path segment by $\beta _{\mathrm{t},l}$, $\beta _{\mathrm{d},k,l}$ and $ {{\bm \psi} }_{k,l} $, respectively.
	Hence, when the UAV is at the $l$th path segment, the expression of the expected received power $ \hat{P}_{k,l} $ at Sensor $ k $ is the same as \eqref{hatP_to_phi_FHB}.
	
	In this section, we propose to provide the required energy for all sensors with minimum energy consumption of the UAV,
	via jointly optimizing the trajectory $\mathbf{q}_{\mathrm{U}}$,
	the flight time $\mathbf{t}$
	and the reflection coefficient vectors $\left\{ {{\bm{\phi}} }_l \right\}$.
	The energy minimization problem is formulated as follows
	\begin{subequations}\label{Prob1}
		\begin{alignat}{2}
			\min_{{\mathbf{q}_{\mathrm{U}},\mathbf{t},\left\{ {{\bm{\phi}} }_l \right\} }} \quad
			& E_{\mathrm{U}}\left( \mathbf{q}_{\mathrm{U}},\mathbf{t} \right) \\
			\mbox{s.t.}\qquad
			&\delta_l\leqslant \min \left\{ \varDelta _{\max},V_{\max}t_l \right\} ,\;l \in \mathcal{L} _{\mathrm{a}},\label{constraint_SegLength}\\
			&t_l\geqslant 0, \;l \in \mathcal{L} _{\mathrm{a}}, \label{constraint_t}\\
			&\eta \sum_{l \in \mathcal{L} _{\mathrm{a}}}{t_l \hat P_{k,l}}\geqslant {E_{k}^{\mathrm{req}}}, \;k=1,\dots ,K,\label{constraint_E_Prob1}\\
			&\eqref{qI&qF}, \eqref{constraint_theta_PD}.\notag
		\end{alignat}
	\end{subequations}
	The main notations used in Problem \eqref{Prob1} are summarized in Table \ref{Table_notation}.
	Note that in Problem \eqref{Prob1}, the objective function $ E_{\mathrm{U}}\left( \mathbf{q}_{\mathrm{U}},\mathbf{t} \right) $ and the constraints \eqref{constraint_theta_PD} and \eqref{constraint_E_Prob1} are non-convex.
	Moreover, $ E_{\mathrm{U}}\left( \mathbf{q}_{\mathrm{U}},\mathbf{t} \right) $ and constraint \eqref{constraint_E_Prob1} are not tractable due to the highly coupled variables.
	Fortunately, Problem \eqref{Prob1} has a similar form to Problem \eqref{Prob1_FHB}.
	By extending the method in Section \ref{FLY-HOVER-BROADCAST PROTOCOL}, we propose an efficient algorithm for solving Problem \eqref{Prob1} in the following.

	\subsection{Optimizing the Trajectory $\mathbf{q}_{\mathrm{U}}$ and Flight Time $\mathbf{t}$}
	The proposed method for joint optimization of $\mathbf{q}_{\mathrm{U}}$ and $\mathbf{t}$ under the FHB protocol in Section \ref{section_opt_qU_t_FHB} provides an efficient algorithm framework for the PD protocol.
	By substituting \eqref{P_p2} into \eqref{E_U}, we can rewrite $E_{\mathrm{U}}\left( \mathbf{q}_{\mathrm{U}},\mathbf{t} \right)$ as follows 
	\begin{align}\label{E_U_full}
		E_{\mathrm{U}}\left( \mathbf{q}_{\mathrm{U}},\mathbf{t} \right) &=\sum_{l \in \mathcal{L} _{\mathrm{a}}}{P_{\mathrm{t}}t_l}+\sum_{l \in \mathcal{L} _{\mathrm{a}}}{P_{\mathrm{i}}\left( \sqrt{t_{l}^{4}+\frac{\delta _{l}^{4}}{4v_{0}^{4}}}-\frac{\delta _{l}^{2}}{2v_{0}^{2}} \right) ^{\frac{1}{2}}} \notag \\
		&\quad+\sum_{l \in \mathcal{L} _{\mathrm{a}}}{P_0\left( t_l+\frac{3\delta _{l}^{2}}{U_{\mathrm{tip}}^{2}t_l} \right) }
		+\frac{1}{2}d_0\rho sA\sum_{l \in \mathcal{L} _{\mathrm{a}}}{\frac{\delta _{l}^{3}}{t_{l}^{2}}}.
	\end{align}
	Then, the subproblem for jointly optimizing the trajectory $\mathbf{q}_{\mathrm{U}}$ and the flight time $\mathbf{t}$ under the PD protocol is formulated as follows
	\begin{subequations}\label{ProbU}
		\begin{alignat}{2}
			\min_{{\mathbf{q}_{\mathrm{U}},\mathbf{t} }} \quad
			& E_{\mathrm{U}}\left( \mathbf{q}_{\mathrm{U}},\mathbf{t} \right) \\
			\mbox{s.t.}\quad
			&\eqref{qI&qF},\eqref{constraint_SegLength}-\eqref{constraint_E_Prob1}.\notag
		\end{alignat}
	\end{subequations}
	Compared with the objective function of Problem \eqref{ProbU_FHB}, that of Problem \eqref{ProbU} is much more complicated.
	In the following, we derive a lower bound for $ E_{\mathrm{U}}\left( \mathbf{q}_{\mathrm{U}},\mathbf{t} \right) $ by applying the SCA method.
	To begin with, we introduce a set of slack variables $\left\{ x_l\geqslant 0 \right\} $ and let
	\begin{equation}\label{x_define}
		x_{l}^{2}\geqslant \sqrt{t_{l}^{4}+\frac{\delta _{l}^{4}}{4v_{0}^{4}}}-\frac{\delta _{l}^{2}}{2v_{0}^{2}}, \;l \in \mathcal{L} _{\mathrm{a}}.
	\end{equation}
	It can be derived that
	\begin{equation}\label{x_derive}
		\frac{t_{l}^{4}}{x_{l}^{2}}\leqslant x_{l}^{2}+\frac{\delta _{l}^{2}}{v_{0}^{2}}, \;l \in \mathcal{L} _{\mathrm{a}}.
	\end{equation}
	By taking the first-order Taylor expansion of the right hand side of \eqref{x_derive} w.r.t. $ x $ and $ \delta $, we have the following inequality
	\begin{equation}\label{x_SCA}
		x_{l}^{2}+\frac{\delta _{l}^{2}}{v_{0}^{2}} \geqslant 2x_{l}^{n}x_l-x_{l}^{2,n}+\frac{2}{v_{0}^{2}}\delta _{l}^{n}\delta _l-\frac{1}{v_{0}^{2}}\delta _{l}^{2,n}, \;l \in \mathcal{L} _{\mathrm{a}},
	\end{equation}
	where $x_{l}^{n}$ and $ \delta _{l}^{n} $ are the value of $x_{l}$ and $ \delta _{l} $ at the $n$th iteration, respectively.
	Then, to deal with the last term on the right hand side of \eqref{E_U_full}, we introduce three sets of slack variables $\left\{ \bar{\delta}_l \right\} $ $\left\{ w_l \right\} $ and $\left\{ z_l \right\} $ with
	\begin{equation}\label{bar_delta_define}
		\delta _l \leqslant \bar{\delta}_l, \;l \in \mathcal{L} _{\mathrm{a}},
	\end{equation}
	\begin{equation}\label{z_define}
		\frac{\bar{\delta}_{l}^{3}}{t_{l}^{2}} \leqslant \frac{z_{l}^{2}}{\bar{\delta}_l}, \;l \in \mathcal{L} _{\mathrm{a}},
	\end{equation}	
	\begin{equation}\label{w_define}
		\frac{z_{l}^{2}}{\bar{\delta}_l}\leqslant w_l, \;l \in \mathcal{L} _{\mathrm{a}}.
	\end{equation}
	Denote by $ z_{l}^{n} $ the value of $ z_{l} $ at the $ n $th iteration.
	By utilizing the first-order Taylor expansion, we approximately transform the inequalities in \eqref{z_define} to
	\begin{equation}\label{z_derive}
		\frac{\bar{\delta}_{l}^{4}}{t_{l}^{2}}\leqslant 2z_{l}^{n}z_l-z_{l}^{2,n}, \;l \in \mathcal{L} _{\mathrm{a}}.
	\end{equation}
	Then, a lower bound for $ E_{\mathrm{U}}\left( \mathbf{q}_{\mathrm{U}},\mathbf{t} \right) $ is obtained as follows
	\begin{align}\label{E_U_lowerbound}
		\hat{E}_{\mathrm{U}}\left( \mathbf{q}_{\mathrm{U}},\mathbf{t} \right) 
		&=\sum_{l\in \mathcal{L} _{\mathrm{a}}}{P_{\mathrm{t}}t_l}+\sum_{l\in \mathcal{L} _{\mathrm{a}}}{P_{\mathrm{i}}x_l}+\sum_{l\in \mathcal{L} _{\mathrm{a}}}{P_0\left( t_l+\frac{3\delta _{l}^{2}}{U_{\mathrm{tip}}^{2}t_l} \right)}  \notag \\
		&\quad+\frac{1}{2}d_0\rho sA\sum_{l\in \mathcal{L} _{\mathrm{a}}}{w_l}.
	\end{align}
	By replacing $ E_{\mathrm{U}}\left( \mathbf{q}_{\mathrm{U}},\mathbf{t} \right)  $ with $ \hat{E}_{\mathrm{U}}\left( \mathbf{q}_{\mathrm{U}},\mathbf{t} \right)  $, we can formulate the problem at the $n$th SCA iteration as follows
	\begin{subequations}\label{ProbU_SCA1}
		\begin{alignat}{2}
			\!\!\!\min_{\mathbf{q}_{\mathrm{U}},\mathbf{t},\{ x_l \},
				\atop
				\{\bar{\delta}_l \} ,\{ w_l \} ,\{ z_l \}} \quad
			& \hat{E}_{\mathrm{U}}\left( \mathbf{q}_{\mathrm{U}},\mathbf{t} \right) \\
			\mbox{s.t.}\qquad\quad
			& \frac{t_{l}^{4}}{x_{l}^{2}}\leqslant 2x_{l}^{n}x_l-x_{l}^{2,n}+\frac{2}{v_{0}^{2}}\delta _{l}^{n}\delta _l-\frac{1}{v_{0}^{2}}\delta _{l}^{2,n}, \notag\\
			&\;l \in \mathcal{L} _{\mathrm{a}},\label{constraint_x}\\
			&\eqref{qI&qF},\eqref{constraint_SegLength}-\eqref{constraint_E_Prob1},\eqref{bar_delta_define},\eqref{w_define},\eqref{z_derive}.\notag
		\end{alignat}
	\end{subequations}
	Note that in Problem \eqref{ProbU_SCA1}, constraint \eqref{constraint_E_Prob1} is still non-convex.
	We follow the similar strategy as in Section \ref{section_opt_qU_t_FHB} to address this energy constraint.
	By taking the approximation in \eqref{hatP_approx}, we first transform constraint \eqref{constraint_E_Prob1} as follows
	\begin{align}\label{constraint_E_approx}
		&\frac{\eta}{E_{k}^{\mathrm{req}}}\sum_{l \in \mathcal{L} _{\mathrm{a}}}P_{\mathrm{t}}t_l\big( \left( U_{1,k,l}^n+U_{3,k,l} \right) \beta _{\mathrm{t},l}+U_{2,k,l}^n\sqrt{\beta _{\mathrm{d},k,l}\beta _{\mathrm{t},l}} \notag \\
		&+\beta _{\mathrm{d},k,l} \big)\ge 1, \;k=1,\dots, K,
	\end{align}
	where $U_{1,k,l}^n$, $U_{2,k,l}^n$ and $U_{3,k,l}$ at the $n$th iteration are given in \eqref{U1}-\eqref{U3}, respectively.
	In addition, it can be readily verified that the inequalities in \eqref{bar_beta_d} and \eqref{bar_beta_t} still hold for $ \beta _{\mathrm{t},l} $ and $ \beta _{\mathrm{d},k,l} $ under the PD protocol.
	Thus, \eqref{constraint_E_approx} is replaced with the following constraints
	\begin{align}\label{constraint_E_ProbU_SCA1}
		&\frac{\eta}{E_{k}^{\mathrm{req}}}\sum_{l \in \mathcal{L} _{\mathrm{a}}}P_{\mathrm{t}}t_l\left(  \left( U_{1,k,l}^n+U_{3,k,l} \right) y_{\mathrm{t},l} + U_{2,k,l}^n\sqrt{y_{\mathrm{t},l}y_{\mathrm{d},k,l}} \right. \notag \\
		&\left. +y_{\mathrm{d},k,l} \right) \ge 1,  \;k=1,\dots ,K,
	\end{align}
	\begin{equation}\label{constraint_yt_ProbU_SCA}
		\bar{\beta}_{\mathrm{t},l}\geqslant y_{\mathrm{t},l}\geqslant 0,  \;l \in \mathcal{L} _{\mathrm{a}},
	\end{equation}
	\begin{equation}\label{constraint_yd_ProbU_SCA}
		\bar{\beta}_{\mathrm{d},k,l}\geqslant y_{\mathrm{d},k,l}\geqslant 0, \;k=1,\dots ,K,
	\end{equation}
	where $\left\{ y_{\mathrm{t},l}\right\} $ and $\left\{ y_{\mathrm{d},k,l}\right\}$ are slack variables.
	Furthermore, by introducing slack variables $\left\{ y_{\mathrm{a},k,l} \right\}$ and $\left\{ e_{k,l} \right\}$, we transform \eqref{constraint_E_ProbU_SCA1} into the following constraints
	\begin{align}\label{constraint_E_ProbU_SCA2}
		&\left( U_{1,k,l}^n+U_{3,k,l} \right) y_{\mathrm{t},l}+U_{2,k,l}^ny_{\mathrm{a},k,l}+y_{\mathrm{d},k,l}\geqslant \frac{E_{k}^{\mathrm{req}}e_{k,l}^{2}}{\eta P_{\mathrm{t}}t_l},  \notag \\
		&\;{l \in \mathcal{L} _{\mathrm{a}}} ,\;k=1,\dots ,K,
	\end{align}
	\begin{equation}\label{constraint_ya_ProbU_SCA}
		y_{\mathrm{t},l}\geqslant \frac{y_{a,k,l}^{2}}{y_{\mathrm{d},k,l}}, \;{l \in \mathcal{L} _{\mathrm{a}}},\;k=1,\dots ,K,
	\end{equation}
	\begin{equation}\label{constraint_e_ProbU_SCA}
		\sum_{l \in \mathcal{L} _{\mathrm{a}}}{\left( 2e_{k,l}^ne_{k,l}-e_{k,l}^{2,n} \right)}\geqslant 1,\;k=1,\dots ,K,
	\end{equation}
	where $e_{k,l}^n$ is the value of $e_{k,l}$ at the $n$th iteration.
	In \eqref{constraint_e_ProbU_SCA}, we have used the fact that the first-order Taylor expansion is a lower bound of a convex function.
	Finally, Problem \eqref{ProbU_SCA1} is reformulated as follows
	\begin{subequations}\label{ProbU_SCA2}
		\begin{alignat}{2}
			\!\!\!\min_{\mathbf{q}_{\mathrm{U}},\mathbf{t},\{ x_l \},\{\bar{\delta}_l \},\{ w_l \} ,\{ z_l \},
				\atop
				\{ y_{\mathrm{t},l}\},\{ y_{\mathrm{d},k,l}\},\{ y_{\mathrm{a},k,l} \},\{ e_{k,l} \} } \quad
			& \hat{E}_{\mathrm{U}}\left( \mathbf{q}_{\mathrm{U}},\mathbf{t} \right) \\
			\mbox{s.t.}\quad\qquad\qquad\;
			&\eqref{qI&qF},\eqref{constraint_SegLength},\eqref{constraint_t},\eqref{bar_delta_define},\eqref{w_define}, \notag \\
			&\eqref{z_derive},\eqref{constraint_x},\eqref{constraint_yt_ProbU_SCA}-\eqref{constraint_e_ProbU_SCA}.\notag
		\end{alignat}
	\end{subequations}
	Problem \eqref{ProbU_SCA2} is a convex problem that can be efficiently solved by the existing optimization tools, such as CVX.
	
	\subsection{Optimizing the Reflection Coefficient Vectors $\left\{{\bm{\phi}}_l\right\}$}
	Similar to Section \ref{section_opt_phi_FHB}, we take maximizing the minimum charged energy of all sensors as the objective of optimizing $\left\{{\bm{\phi}}_l\right\}$.
	The corresponding optimization problem is given by
	\begin{subequations}\label{ProbR}
		\begin{alignat}{2}
			\max_{{\left\{ {{\bm{\phi}} }_l \right\},\epsilon }} \quad
			& \epsilon \\
			\mbox{s.t.}\quad \;
			&\frac{\eta}{E_{k}^{\mathrm{req}}} \sum_{l \in \mathcal{L} _{\mathrm{a}}}{t_l \hat P_{k,l}}\geqslant \epsilon,\;k=1,\dots ,K,\\
			&\eqref{constraint_theta_PD}.\notag
		\end{alignat}
	\end{subequations}
	Note that Problem \eqref{ProbR} has the same form as Problem \eqref{ProbR_FHB} except for the value range of $ l $. 
	Therefore, by setting $ L $ as the number of path segments and the index of $l$ as $ \mathcal{L}_{\mathrm{a}} $, the proposed Algorithm \ref{Alg_MM} in Section \ref{section_opt_phi_FHB} can be directly applied for solving Problem \eqref{ProbR}.

	\subsection{Algorithm Development}
	\begin{algorithm}[t]
		\caption{SCA-MM algorithm under the PD protocol}
		\label{Alg_SCA-MM_PD}
		\begin{algorithmic}[1] %此处的[1]控制一下算法中的每句前面都有标号
			\Require  Initial the number of iterations as $n=1$. Set feasible ${{\bm{\phi}}}^1$, $\mathbf{q}_{\mathrm{U}}^1$ and $\mathbf{t}^1$, the maximum number of iterations $n_{\max}$, and smoothing-related factors $\mu,\mu _{\max}$ and $\iota$.
			\Repeat
			\State Given $\mathbf{q}_{\mathrm{U}}^n$, calculate $U_{1,k,l}^n$, $U_{2,k,l}^n$ and $U_{3,k,l}$ in \eqref{U1}-\eqref{U3};
			\State Set $\left\{x_{l}^{n}\right\}$, $\left\{z_{l}^{n}\right\}$ and $\{e_{k,l}^n\}$ to hold the equalities in \eqref{x_define}, \eqref{z_define} and \eqref{constraint_E_ProbU_SCA2}, respectively;
			\State Given ${{\bm{\phi}}}^n$, calculate the optimal $\mathbf{q}_{\mathrm{U}}^{n+1}$ and $\mathbf{t}^{ n+1 }$ by solving Problem \eqref{ProbU_SCA2};\label{update_qU_T_SCA-MM_PD}
%			\State Given $\mathbf{q}_{\mathrm{U}}^{n+1}$, update $ {{\bm \psi}}_{k,l} $ in \eqref{Varphi};
			\State Given $\mathbf{q}_{\mathrm{U}}^{n+1}$ and $\mathbf{t}^{ n+1 }$, calculate the optimal ${{\bm{\phi}}}^{ n+1 }$ with Algorithm \ref{Alg_MM};\label{update_phi_SCA-MM_PD}
			\State Set $\mu \gets \max \left\{ \mu ^{\iota},\mu _{\max} \right\} $ and $n\gets n+1$;
			\Until $n\geqslant n_{\max}$.
		\end{algorithmic}
	\end{algorithm}

	Based on the above discussions, we develop the SCA-MM method for solving Problem \eqref{Prob1}, which is summarized in Algorithm \ref{Alg_SCA-MM_PD}.
	It can be found that Algorithm \ref{Alg_SCA-MM_PD} follows a similar framework as Algorithm \ref{Alg_SCA-MM_FHB}, where the optimal ${{\bm{\phi}}}^{ n+1 }$ is calculated with Algorithm \ref{Alg_MM} in Step \ref{update_phi_SCA-MM_PD}.
	Compared with the FHB protocol, the PD protocol achieves higher flexibility in trajectory and flight time planning, which further reduces the energy consumption of the UAV.
	
		Similar to Algorithm \ref{Alg_SCA-MM_FHB}, the computational complexity of Algorithm \ref{Alg_SCA-MM_PD} mainly lies in solving Problem \eqref{ProbU_SCA2} in Step \ref{update_qU_T_SCA-MM_PD} and calculating ${{\bm{\phi}}}^{ n+1 }$ in Step \ref{update_phi_SCA-MM_PD}.
		Specifically, the complexity order of solving Problem \eqref{ProbU_SCA2} is $\mathcal{O}^{\mathrm{Alg. 3}}_{\mathrm{S}4}=\mathcal{O} ( \sqrt{5KL+2K+11L}( n^2KL^2+n KL^3 +n^3 ) ) $, where $n= KL$.
		In addition, the complexity of calculating ${{\bm{\phi}}}^{ n+1 }$ is of order $\mathcal{O}^{\mathrm{Alg. 3}}_{\mathrm{S}5}=\mathcal{O} ( r_{\mathrm{MM}}K( L^2M^2+M^3 ) ) $.
		Hence, the overall complexity of Algorithm \ref{Alg_SCA-MM_PD} is $\mathcal{O}^{\mathrm{Alg. 3}}_{\mathrm{S}4}+\mathcal{O}^{\mathrm{Alg. 3}}_{\mathrm{S}5}$.

	\section{Simulation Results} \label{SIMULATION RESULTS}
	In this section, extensive simulation results are presented to investigate the performance of the RIS-assisted UAV-enabled WPT system with the proposed algorithms.

	\begin{figure}[t]
		\centering
		\includegraphics[width=0.9\linewidth]{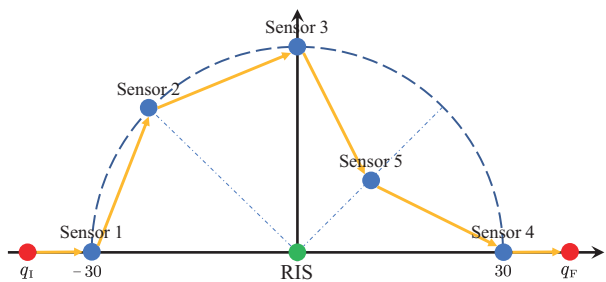}
		\caption{The simulated RIS-assisted UAV-enabled WPT scenario and the initial trajectory of the UAV.}
		\label{Figsimulmodel}
	\end{figure}

	\subsection{Simulation Setup}
	Fig. \ref{Figsimulmodel} illustrates the simulation scenario, where $ K=5 $ sensors are considered to be distributed in a semicircular area with a radius of 30 m.
	As shown in Fig. \ref{Figsimulmodel}, the semicircle is divided into four equal parts by five radii, and there is a sensor on each radius.
	Specifically, Sensor 1-4 are located on the circle, while Sensor 5 is located at the midpoint of the radius.
	The RIS is deployed at the center of the semicircle with a height of $ H_R=10 $ m.
	The flight height, maximum speed and radiated power of the UAV are set as $ H_U=20 $ m, $V_{\max}=30$ m/s and $ P_{\mathrm t}=40 $ dBm, respectively.
	The initial and final locations are $q_{\mathrm{I}}=-35+j0$ and $q_{\mathrm{F}}=35+j0$.
	The parameters associated with UAV's propulsion power are set to the same as in \cite[Table 1]{8663615}, resulting in the MR speed of ${v_{\mathrm{mr}}}=18.3$ m/s.
	Unless otherwise specified, the simulation parameters are set as follows: 
	the energy requirement for each sensor of $ E_{k}^{\mathrm{req}}=E^{\mathrm{req}}=0.2 $ mJ, path loss exponents of $ \alpha _{\mathrm{t}} = \alpha _{\mathrm{r}} =2.2$ and $ \alpha _{\mathrm{d}}=2.6 $, Rician factors of $\kappa _{\mathrm{t}}=\kappa _{\mathrm{r}}=\kappa _{\mathrm{d}}=10$, $ \lambda=1 $ m, $ d=0.5 $ m, $ \eta=0.6 $, $ \varDelta _{\max}=0.5 $ m,
	$\varepsilon _{\mathrm{e}}=10^{-6}$, $ r_{\max}=10 $, $ \mu=100 $, $ \mu _{\max}=1000 $, $ \iota=1.07 $, $ n _{\max}=60 $.
	
	In the following simulation results, the scheme with Algorithm \ref{Alg_SCA-MM_FHB} under the FHB protocol is denoted by \textbf{FHB}, and that with Algorithm \ref{Alg_SCA-MM_PD} under the PD protocol is denoted by \textbf{PD}.
	For both protocols, the reflection coefficient vectors $\left\{ {{\bm{\phi}} }_l \right\}$ are initialized by setting $\theta _{m,l}=0$, $\forall m,l$.
	In \textbf{FHB}, 5 hovering locations ($ L=6 $) are set with the same plane coordinates as the sensors, and the initialized trajectory $ \mathbf{q}_{\mathrm{U}} $ is illustrated by the yellow arrows in Fig. \ref{Figsimulmodel}.
	In \textbf{PD}, the initial value of $ \mathbf{q}_{\mathrm{U}} $ is obtained by equally dividing each initialized path segment of \textbf{FHB} into portions with length not exceeding $\frac{\varDelta _{\max}}{1.8}$ ($ L=362 $),
	and $ \mathbf t $ is initialized by setting $ v_l=v_{\mathrm{mr}}$ for $ l \in \mathcal{L} _{\mathrm{a}} $.

	\subsection{Convergence of Proposed Algorithms}
\begin{figure}
		\centering
		\includegraphics[width = 0.9\linewidth]{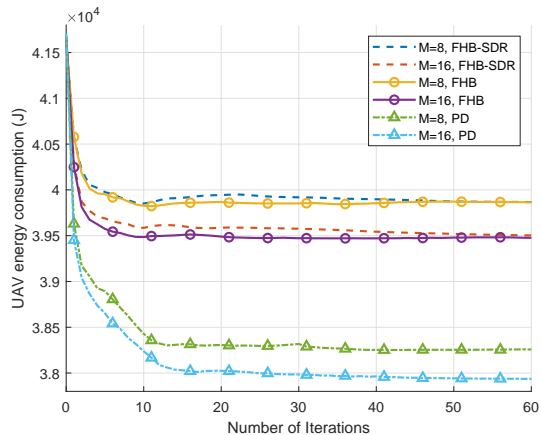}
		\caption{Convergence of proposed algorithms for $M=\left[8,16\right]$.}
		\label{fig:convergence}
	\end{figure}

	We first study the convergence of our proposed algorithms.
	For comparison, the following baseline schemes are also investigated:
	\begin{enumerate}
		\item \textbf{FHB-SDR}: To analyze the performance of our proposed Algorithm \ref{Alg_MM}, this scheme solves Problem \eqref{ProbR_FHB} in \textbf{PD} by SDR-based method with MOSEK solver \cite{MOSEKtoolbox} and $ 10^{4} $ Gaussian randomization operations.
		
		\item \textbf{FHB-noRIS} and \textbf{PD-noRIS}: In these two schemes with no RIS deployed, the expected received power is given by $\hat{P}_{k,l}^{\mathrm{noRIS}}=\mathbb{E} \left\{ P_{\mathrm{t}}\left| g_{\mathrm{d},k,l} \right|^2 \right\} =P_{\mathrm{t}}\beta _{\mathrm{d},k,l}$.
		The corresponding energy minimization problems are solved by the methods we propose to solve Problem \eqref{ProbU_FHB} and \eqref{ProbU}.
		
		\item \textbf{FHB-2bit} and \textbf{PD-2bit}: Due to hardware limitations, continuously tunable phase shifts are impractical to be implemented on the RIS.
		To investigate the system performance in practical scenarios, these two schemes with 2 bit resolution are considered,
		where each element of the optimal phase shifts $ \left\{ \theta _{m,l}\right\} $ obtained by Algorithm \ref{Alg_SCA-MM_FHB} and Algorithm \ref{Alg_SCA-MM_PD} is quantized  as $ 0 $, $ \frac{\pi }{2} $, $ \pi $ or $ \frac{{3\pi }}{2} $,
%		and $ \mathbf{q}_{\mathrm{U}} $ and $ \mathbf{t} $ are updated accordingly.
%		where each element of the optimal phase shifts $ \left\{ \theta _{m,l}\right\} $ of Algorithm \ref{Alg_SCA-MM_FHB} and Algorithm \ref{Alg_SCA-MM_PD} is quantized  as $ 0 $, $ \frac{\pi }{2} $, $ \pi $ or $ \frac{{3\pi }}{2} $,
		and $ \mathbf{q}_{\mathrm{U}} $ and $ \mathbf{t} $ are updated accordingly.
	\end{enumerate}
	Fig. \ref{fig:convergence} plots the energy consumption of the UAV versus the number of iterations.
	It can be observed that \textbf{FHB} and \textbf{PD} converge in a few iterations.
	Additionally, thanks to the update strategies in Algorithm \ref{Alg_SCA-MM_FHB} and Algorithm \ref{Alg_SCA-MM_PD}, there is no obvious fluctuation.
	These demonstrate the effectiveness of our proposed algorithms in the joint optimization of the UAV's trajectory, flight time and the RIS's reflection coefficients.
	Moreover, Fig. \ref{fig:convergence} shows that \textbf{FHB} outperforms \textbf{FHB-SDR} in both performance and convergence speed, which validates the advantages of Algorithm \ref{Alg_MM} in the passive beamforming design of the RIS for the entire flight.

	%	\begin{figure}[!htb]
	%		\begin{tabular}{cc}
	%			\begin{minipage}[t]{0.48\linewidth}
	%				\includegraphics[width = 1\linewidth]{convergence.eps}
	%				\caption{Convergence of proposed algorithms for $M=\left[8,16\right]$.}
	%				\label{fig:convergence}
	%			\end{minipage}
	%			\begin{minipage}[t]{0.48\linewidth}
	%				\includegraphics[width = 1\linewidth]{M.eps}
	%				\caption{UAV energy consumption versus the number of the RIS reflecting elements.}
	%				\label{fig:M}
	%			\end{minipage}
	%		\end{tabular}
	%	\end{figure}
	
	\begin{figure}
		\centering
		\includegraphics[width = 0.9\linewidth]{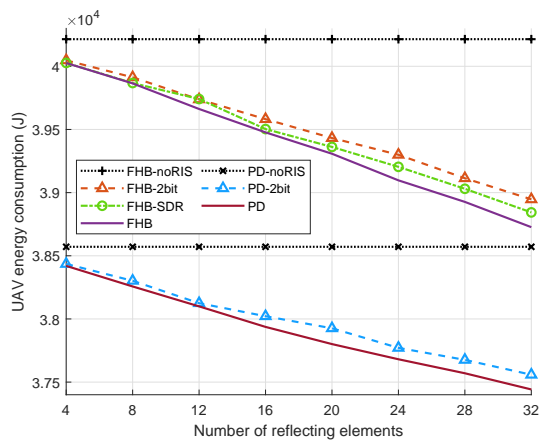}
		\caption{UAV energy consumption versus the number of RIS reflecting elements.}
		\label{fig:M}
	\end{figure}

	\subsection{Energy Saving Performance of RIS}
	Fig. \ref{fig:M} shows the energy consumption of the UAV versus the number of RIS reflecting elements.	
	Firstly, it is observed that the UAV energy consumption under the PD protocol is significantly lower than that under the FHB protocol.
	Due to the excellent flexibility in the planning of the UAV's trajectory and flight time, the optimized solutions of PD protocol achieve superior energy saving performance.
%	This provides a tradeoff for practical system planning, i.e. the cost of computation and control can be traded for the energy saving of the UAV.
%	Therefore, the PD protocol is superior in scenarios with higher computing power and more precise UAV positioning, while the FHB protocol is the opposite.
	Secondly, Fig. \ref{fig:M} shows that the energy consumption of the UAV decreases approximately linearly with the increase of $ M $, 
	which validates the efficiency of RIS on the energy saving.
	In addition, the performance comparison between \textbf{FHB} and \textbf{FHB-SDR} again verifies the advantages of Algorithm \ref{Alg_MM} over SDR-based method.
	Compared with continuous phase shift schemes, it can also be observed that the performance loss of \textbf{2 bit} schemes is slight.

	\subsection{Trajectory and Flight Time Planning of UAV}
	\begin{figure}
		\centering
		\subfigure[$ E^{\mathrm{req}}=0.2 $ mJ]{
			\label{fig:qU_Ereq02}
			\includegraphics[width=0.9\linewidth]{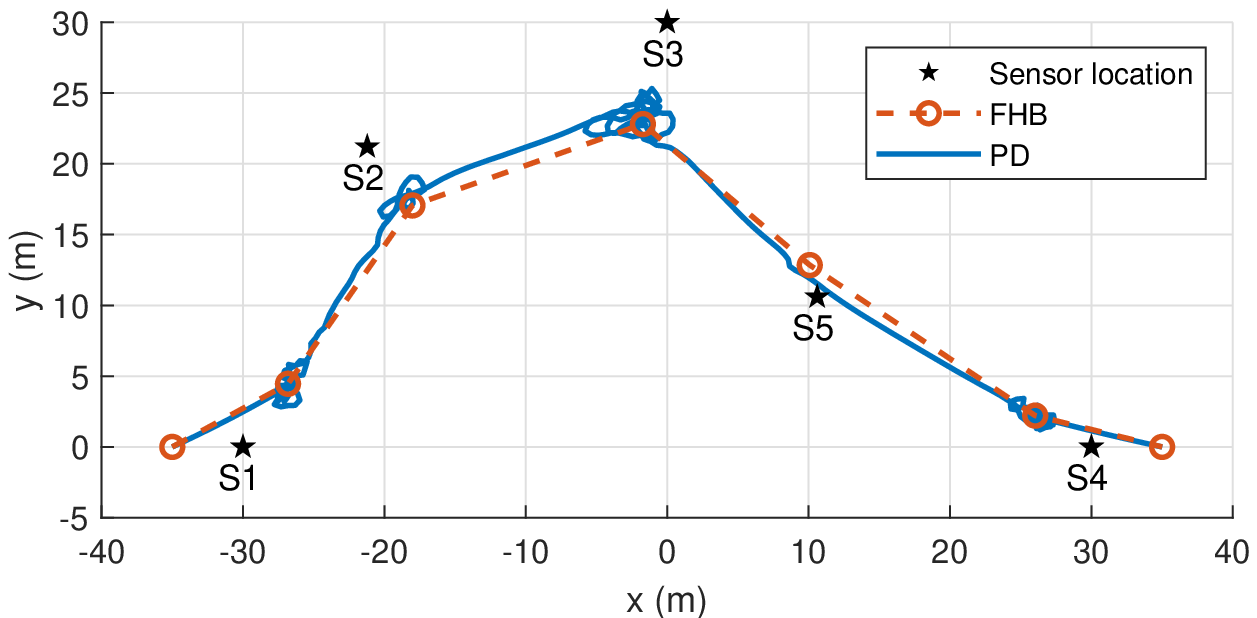}}
		\subfigure[$ E^{\mathrm{req}}=0.02 $ mJ]{
			\label{fig:qU_Ereq002}
			\includegraphics[width=0.9\linewidth]{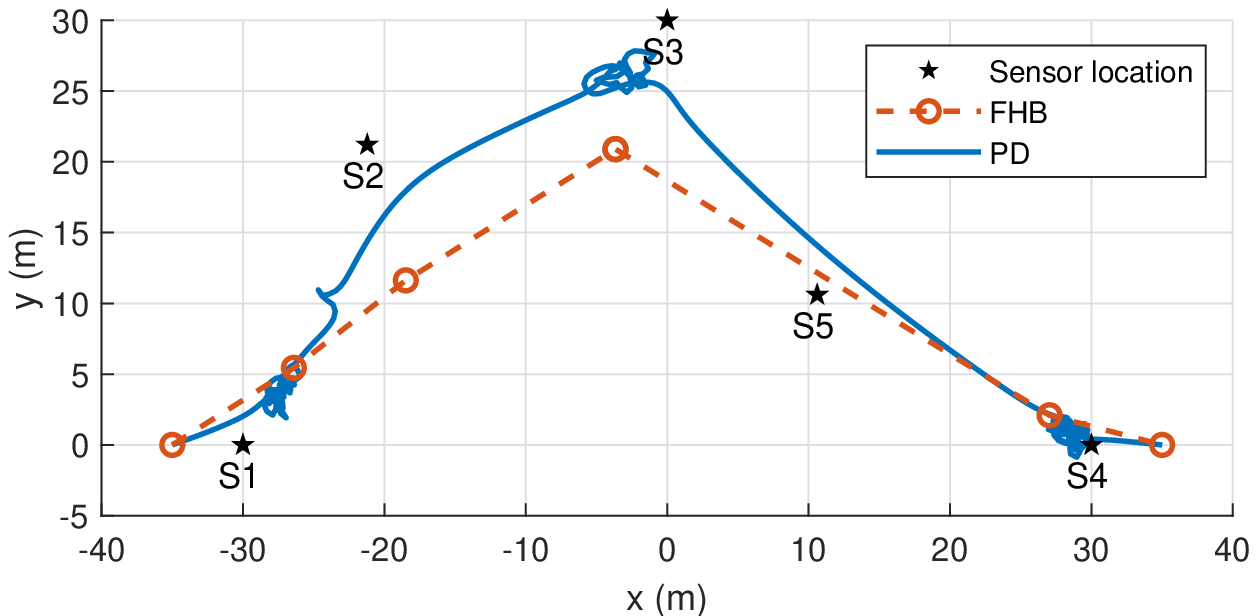}}
		\caption{Optimized UAV's trajectories for different energy requirements of the sensors.}
		\label{fig:qU} %% label for entire figure
	\end{figure}
	\begin{figure}
		\centering
		\subfigure[$ E^{\mathrm{req}}=0.2 $ mJ]{
			\label{fig:vT_Ereq02}
			\includegraphics[width=0.9\linewidth]{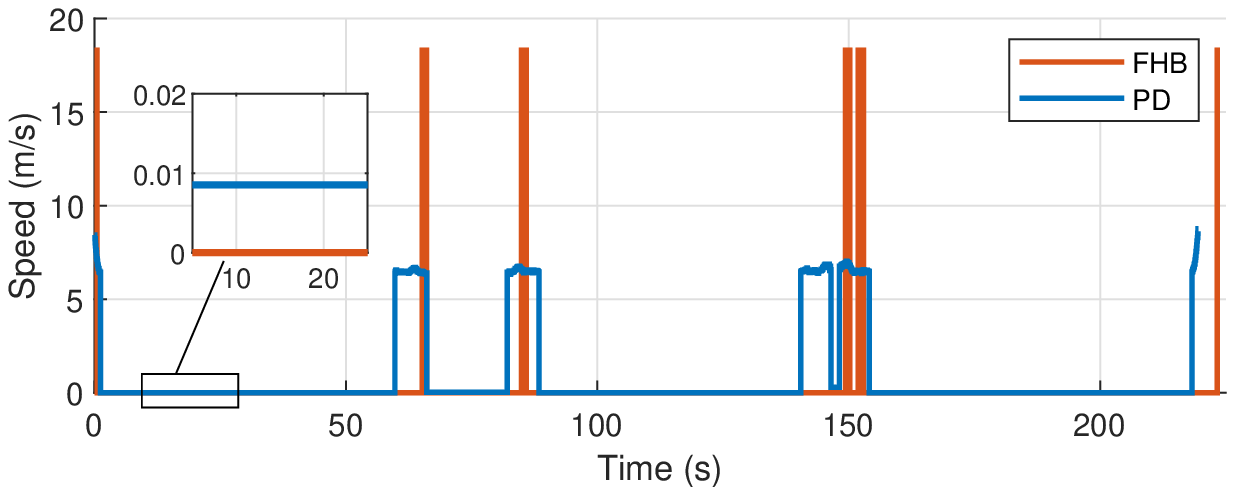}}
		\subfigure[$ E^{\mathrm{req}}=0.02 $ mJ]{
			\label{fig:vT_Ereq002}
			\includegraphics[width=0.9\linewidth]{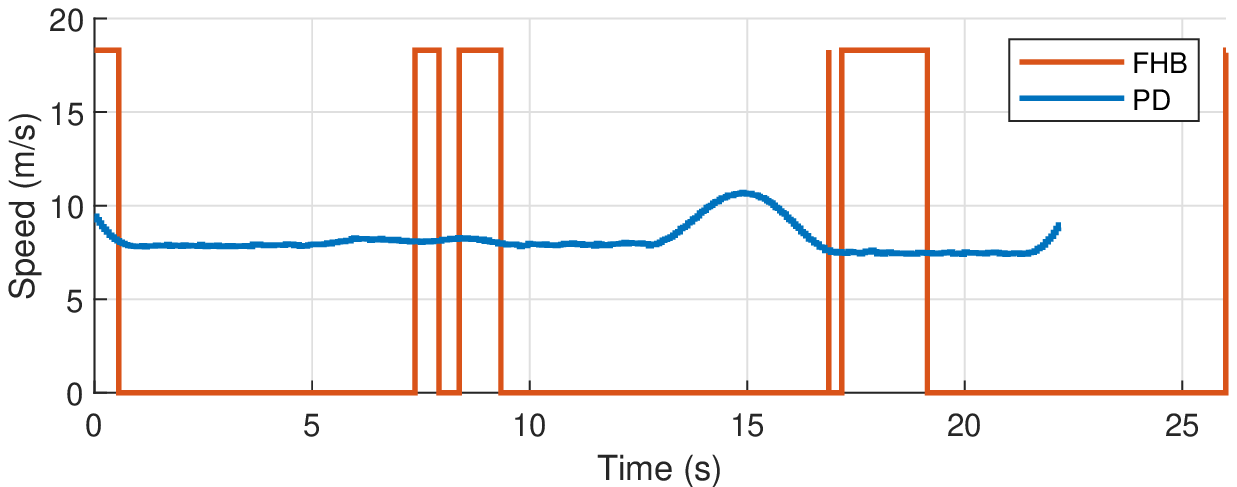}}
		\caption{UAV Speed corresponding to the UAV's trajectories in Fig. \ref{fig:qU}.}
		\label{fig:vT} %% label for entire figure
	\end{figure}
	\begin{figure}
		\centering
		\subfigure[$ E^{\mathrm{req}}=0.2 $ mJ]{
			\label{fig:Phat_Ereq02}
			\includegraphics[width=0.9\linewidth]{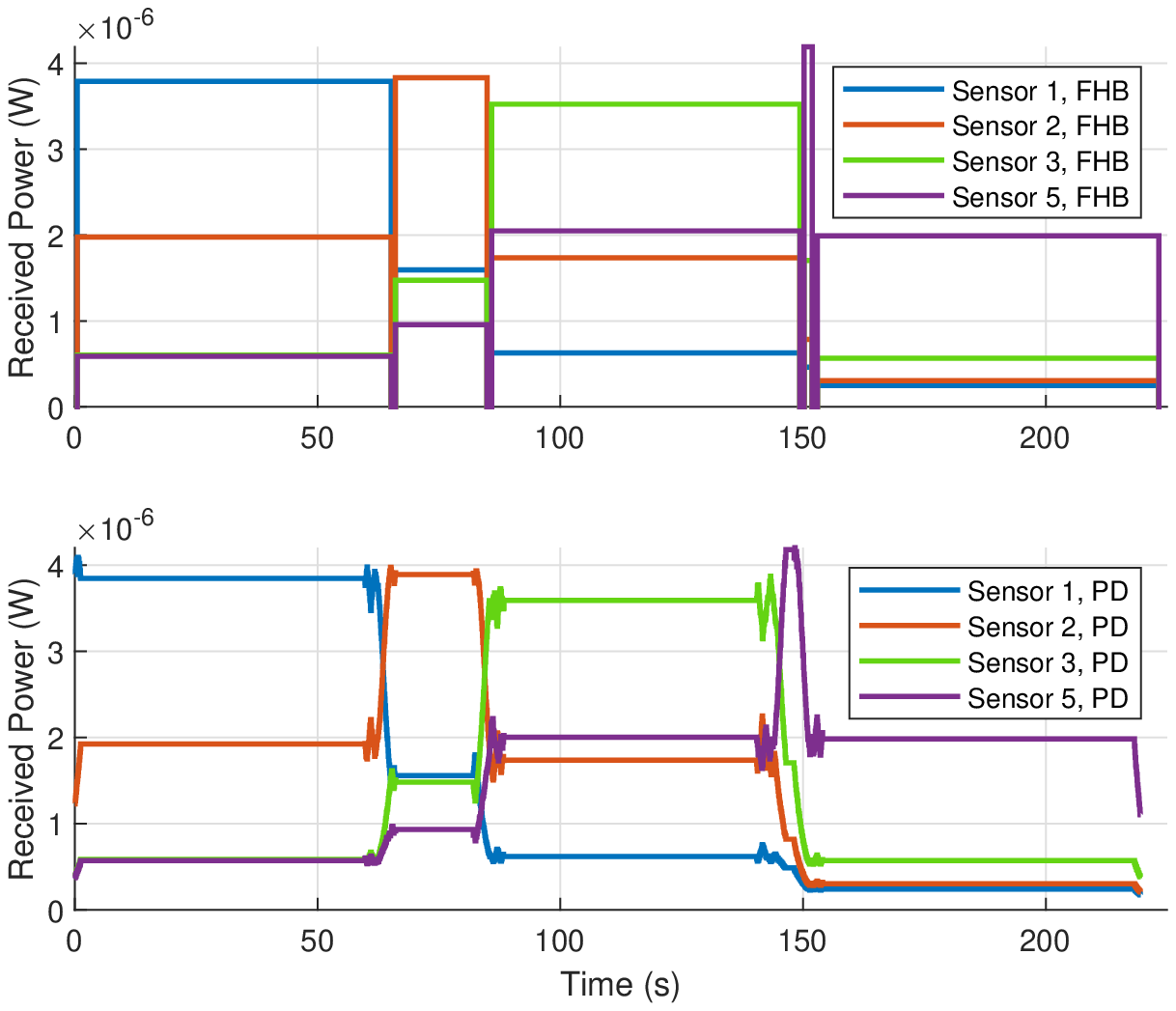}}
		\subfigure[$ E^{\mathrm{req}}=0.02 $ mJ]{
			\label{fig:Phat_Ereq002}
			\includegraphics[width=0.9\linewidth]{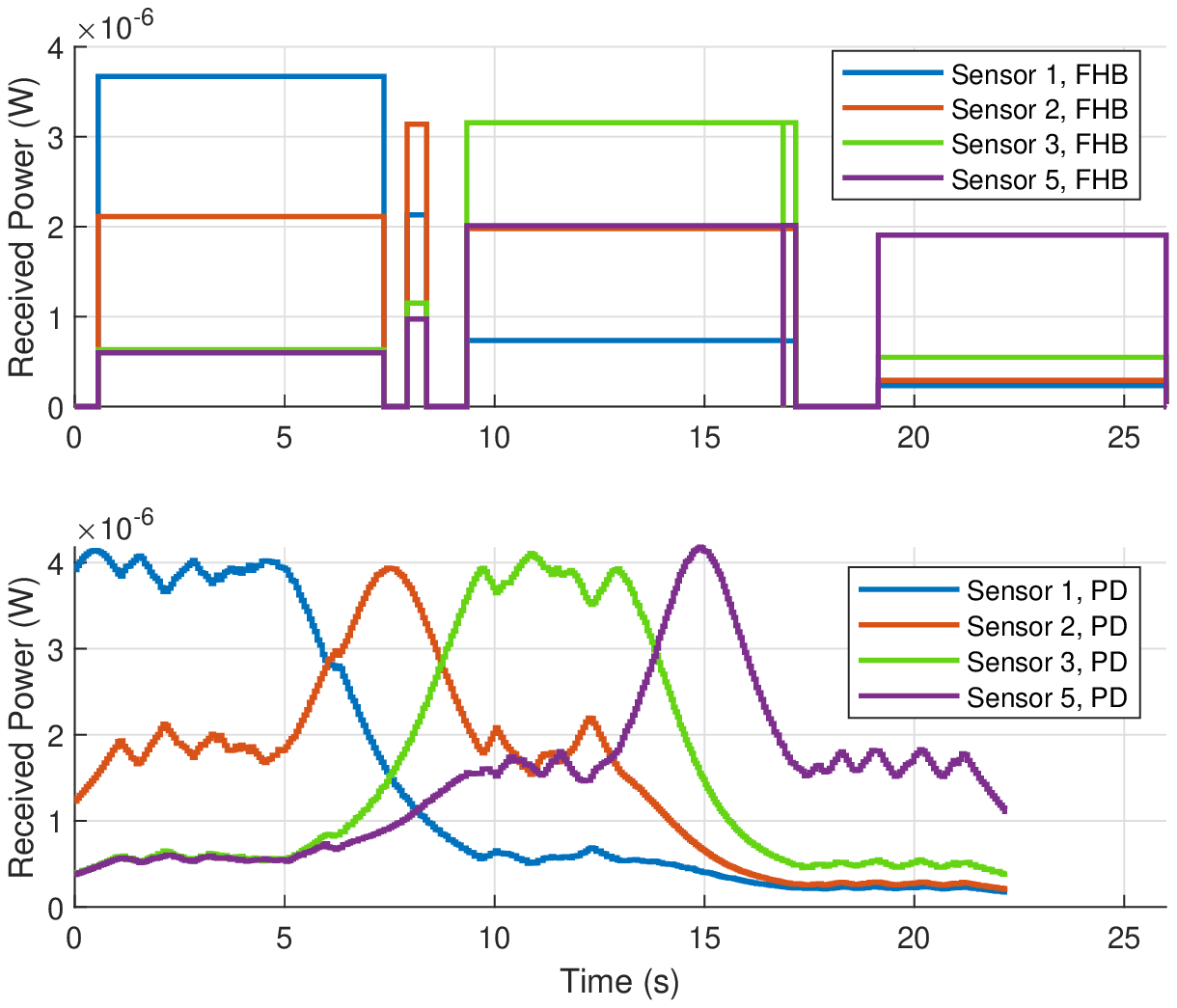}}
		\caption{Received power at Sensor 1, 2, 3 and 5 corresponding to the UAV's trajectories in Fig. \ref{fig:qU}.}
		\label{fig:Phat} %% label for entire figure
	\end{figure}
%	\vspace{-0.2 cm}
	
	For two cases of the energy requirement $ E_{k}^{\mathrm{req}}=0.2 $ mJ and $ E_{k}^{\mathrm{req}}=0.02 $ mJ, Fig. \ref{fig:qU} and Fig. \ref{fig:vT} illustrate the optimized UAV's trajectory and flight time, respectively.
	Firstly, it can be observed from Fig. \ref{fig:qU} that for both protocols, the UAV does not fly directly above the sensors.
	This is as expected since all sensors continually receive RF signals from the UAV and RIS during the flight.
	Radiating directly above each sensor leads to a longer flight distance, and thus increases the power transmission time.
	As we mentioned in Section \ref{section_sysmodel_FHB}, hovering is not the most energy efficient state for a rotor-wing UAV.
	Therefore, Fig. \ref{fig:vT} shows that under the PD protocol, the UAV maintains a certain speed when it reaches an area favorable for power transmission.
	Additionally, in the scenario with lower sensor energy requirements, the UAV speed increases as the time required to radiate decreases, as expected.
	
	In Fig. \ref{fig:Phat}, we compare the received power at Sensor 1, 2, 3 and 5.
%	It is first observed that for the proposed WPT system, 
	Under the PD protocol, the received power at each sensor is larger than 0 regardless of the UAV's location as expected, which maximizes the harvested energy.
	Furthermore, by allowing the UAV to radiate RF signals during its flight, the PD protocol enables greater flexibility for the joint design of UAV's trajectory and flight time.
	From Fig. \ref{fig:qU_Ereq02} and \ref{fig:qU_Ereq002}, \textbf{FHB} tends to reduce both the flight distance and the hovering time as the energy requirements decrease. 
	For the PD protocol, its trend is interesting. 
	From Fig. \ref{fig:Phat_Ereq002}, when time duration is from 0 s to 5 s, the received power for Sensor 1 almost keeps stable at its largest value of $ 4\times10^{-6} $, which means the UAV may hover over Sensor 1 in this time duration. 
	However, for the time duration from 5 s to 10 s, the received power for Sensor 2 firstly increases with the time and then decreases with the time, which implies that the UAV did not hover over Sensor 2. 
	The reason is that Sensor 2 has already harvested enough energy as the UAV keeps transmitting the wireless energy during its flight.

	\section{Conclusions} \label{CONCLUSION}
	This paper studied an energy-efficient UAV-enabled WPT system, where the received power of the RF signals at the sensors is enhanced by an RIS.
	We investigated the problems of providing the required energy for all sensors with minimum UAV energy consumption.
	Considering the FHB protocol for a simple scenario, the trajectory and hovering time of the UAV and the reflection coefficients of the RIS are jointly optimized.
	To solve this complex problem, we decoupled the variables by constructing two subproblems.
	By respectively utilizing the MM method and the SCA framework to solve the subproblems, an efficient iterative algorithm was proposed.
	For the general scenario, we formulated the energy minimization problem that jointly optimizes the UAV's trajectory, flight time and the RIS's reflection coefficients by applying the PD protocol.
	A high-quality solution for this more challenging problem was obtained.
	Our simulation results demonstrated the effectiveness of the proposed algorithms and the energy saving performance of RIS in UAV-enabled WPT systems.

	\begin{appendices}
		\section{Proof of Theorem \ref{theorem_hatP}}\label{appendix_hatP}
		By substituting \eqref{g_t}, \eqref{g_r} and \eqref{g_d} into \eqref{P_k,l_origin}, it can be derived that
		\begin{align}\label{hatP_to_a}
			&\mathbb{E} \left\{ \left| g_{\mathrm{d},k,q}+\mathbf{g}_{\mathrm{r},k,q}^{\mathrm{H}}\mathbf{\Phi }_q\mathbf{g}_{\mathrm{t},q} \right|^2 \right\}=\left| a_{0,k,q} \right|^2+\mathbb{E} \left\{ \left| a_{1,k,q} \right|^2 \right\} \notag \\
			&+\mathbb{E} \left\{ \left| a_{2,k,q} \right|^2 \right\} +\mathbb{E} \left\{ \left| a_{3,k,q} \right|^2 \right\} +\mathbb{E} \left\{ \left| a_{4,k,q} \right|^2 \right\},
		\end{align}
		where
		\begin{subequations}
			\begin{align}
				a_{0,k,q}&=\sqrt{\frac{\kappa _{\mathrm{r}}\beta _{\mathrm{r},k}}{\kappa _{\mathrm{r}}+1}}\sqrt{\frac{\kappa _{\mathrm{t}}\beta _{\mathrm{t},q}}{\kappa _{\mathrm{t}}+1}}\left( \mathbf{g}_{\mathrm{r},k}^{\mathrm{LoS}} \right) ^{\mathrm{H}}\mathbf{\Phi }_q\mathbf{g}_{\mathrm{t},q}^{\mathrm{LoS}}\notag \\
				&\quad+\sqrt{\frac{\kappa _{\mathrm{d}}\beta _{\mathrm{d},k,q}}{\kappa _{\mathrm{d}}+1}}g_{\mathrm{d},k,q}^{\mathrm{LoS}},\label{a0_0}\\
				a_{1,k,q}&=\sqrt{\frac{\beta _{\mathrm{d},k,q}}{\kappa _{\mathrm{d}}+1}}g_{\mathrm{d},k,q}^{\mathrm{NLoS}},\\
				a_{2,k,q}&=\sqrt{\frac{\beta _{\mathrm{r},k}}{\kappa _{\mathrm{r}}+1}}\sqrt{\frac{\kappa _{\mathrm{t}}\beta _{\mathrm{t},q}}{\kappa _{\mathrm{t}}+1}}\left( \mathbf{g}_{\mathrm{r},k,q}^{\mathrm{NLoS}} \right) ^{\mathrm{H}}\mathbf{\Phi }_q\mathbf{g}_{\mathrm{t},q}^{\mathrm{LoS}},\\
				a_{3,k,q}&=\sqrt{\frac{\kappa _{\mathrm{r}}\beta _{\mathrm{r},k}}{\kappa _{\mathrm{r}}+1}}\sqrt{\frac{\beta _{\mathrm{t},q}}{\kappa _{\mathrm{t}}+1}}\left( \mathbf{g}_{\mathrm{r},k}^{\mathrm{LoS}} \right) ^{\mathrm{H}}\mathbf{\Phi }_q\mathbf{g}_{\mathrm{t},q}^{\mathrm{NLoS}},\\
				a_{4,k,q}&=\sqrt{\frac{\beta _{\mathrm{r},k}}{\kappa _{\mathrm{r}}+1}}\sqrt{\frac{\beta _{\mathrm{t},q}}{\kappa _{\mathrm{t}}+1}}\left( \mathbf{g}_{\mathrm{r},k,q}^{\mathrm{NLoS}} \right) ^{\mathrm{H}}\mathbf{\Phi }_q\mathbf{g}_{\mathrm{t},q}^{\mathrm{NLoS}}.
			\end{align}
		\end{subequations}
		It is readily to verify that $\mathbb{E} \left\{ \mathbf{g}_{\mathrm{r},k,q}^{\mathrm{NLoS}}\left( \mathbf{g}_{\mathrm{r},k,q}^{\mathrm{NLoS}} \right) ^{\mathrm{H}} \right\} =\mathbf{I}_M$, $\mathbf{\Phi }_{q}^{\mathrm{H}}\mathbf{\Phi }_q=\mathbf{I}_M$ and $\left( \mathbf{g}_{\mathrm{t},q}^{\mathrm{LoS}} \right) ^{\mathrm{H}}\mathbf{g}_{\mathrm{t},q}^{\mathrm{LoS}}=M$.
		Hence, we have
		\begin{align}
			&\mathbb{E} \left\{ \left| a_{2,k,q} \right|^2 \right\}\notag\\
			&=\mathbb{E} \left\{ \frac{\kappa _{\mathrm{t}}\beta _{\mathrm{r},k}\beta _{\mathrm{t},q}}{\left( \kappa _{\mathrm{r}}+1 \right) \left( \kappa _{\mathrm{t}}+1 \right)}\left( \mathbf{g}_{\mathrm{t},q}^{\mathrm{LoS}} \right) ^{\mathrm{H}}\mathbf{\Phi }_{q}^{\mathrm{H}}\mathbf{g}_{\mathrm{r},k,q}^{\mathrm{NLoS}}\left( \mathbf{g}_{\mathrm{r},k,q}^{\mathrm{NLoS}} \right) ^{\mathrm{H}}\mathbf{\Phi }_q\mathbf{g}_{\mathrm{t},q}^{\mathrm{LoS}} \right\}\notag \\
			&=\frac{M\kappa _{\mathrm{t}}\beta _{\mathrm{r},k}\beta _{\mathrm{t},q}}{\left( \kappa _{\mathrm{r}}+1 \right) \left( \kappa _{\mathrm{t}}+1 \right)}.
		\end{align}
		Similarly, it can be derived that $ \mathbb{E} \left\{ \left| a_{1,k,q} \right|^2 \right\} =\frac{\beta _{\mathrm{d},k,q}}{\kappa _{\mathrm{d}}+1} $, 
		$ \mathbb{E}\! \left\{ \left| a_{3,k,q} \right|^2 \right\} \!=\! \frac{M\kappa _{\mathrm{r}}\beta _{\mathrm{r},k}\beta _{\mathrm{t},q}}{\left( \kappa _{\mathrm{r}}+1 \right) \left( \kappa _{\mathrm{t}}+1 \right)} $ and $ \mathbb{E} \!\left\{ \left| a_{4,k,q} \right|^2 \right\} \!=\!\frac{M\beta _{\mathrm{r},k}\beta _{\mathrm{t},q}}{\left( \kappa _{\mathrm{r}}+1 \right) \left( \kappa _{\mathrm{t}}+1 \right)} $.
		Let us rewrite $ a_{0,k,q} $ as shown in \eqref{a0_formulate} on the top of the next page.
		\begin{figure*}%[hb]
			\begin{align}\label{a0_formulate}
				a_{0,k,q}&=\sqrt{\frac{\kappa _{\mathrm{r}}\beta _{\mathrm{r},k}}{\kappa _{\mathrm{r}}+1}}\sqrt{\frac{\kappa _{\mathrm{t}}\beta _{\mathrm{t},q}}{\kappa _{\mathrm{t}}+1}}\left( e^{-j\frac{2\pi d_{\mathrm{r},k}}{\lambda}}\left[ 1,e^{-j\frac{2\pi d}{\lambda}\cos \omega _{\mathrm{r},k}},...,e^{-j\frac{2\pi \left( M-1 \right) d}{\lambda}\cos \omega _{\mathrm{r},k}} \right] ^{\mathrm{T}} \right) ^{\mathrm{H}} \left[ \begin{matrix}
					e^{j\theta _{1,q}}&		&		\\
					&		\ddots&		\\
					&		&		e^{j\theta _{M,q}}\\
				\end{matrix} \right]\notag
				\\
				&\quad \cdot e^{-j\frac{2\pi d_{\mathrm{t},q}}{\lambda}}\left[ 1,e^{-j\frac{2\pi d}{\lambda}\cos \omega _{\mathrm{t},q}},...,e^{-j\frac{2\pi \left( M-1 \right) d}{\lambda}\cos \omega _{\mathrm{t},q}} \right] ^{\mathrm{T}} +\sqrt{\frac{\kappa _{\mathrm{d}}\beta _{\mathrm{d},k,q}}{\kappa _{\mathrm{d}}+1}}\exp \left( -j\frac{2\pi d_{\mathrm{d},k,q}}{\lambda} \right)\notag
				\\
				&=\sqrt{\frac{\kappa _{\mathrm{r}}\kappa _{\mathrm{t}}\beta _{\mathrm{r},k}\beta _{\mathrm{t},q}}{\left( \kappa _{\mathrm{r}}+1 \right) \left( \kappa _{\mathrm{t}}+1 \right)}}\sum_{m=1}^M{ \exp \left( j\left( \frac{2\pi \left( d_{\mathrm{r},k}-d_{\mathrm{t},q} \right)+2\pi d\left( \cos \omega _{\mathrm{r},k}-\cos \omega _{\mathrm{t},q} \right) \left( m-1 \right)}{\lambda}+\theta _{m,q} \right) \right)}\notag
				\\
				&\quad  +\sqrt{\frac{\kappa _{\mathrm{d}}\beta _{\mathrm{d},k,q}}{\kappa _{\mathrm{d}}+1}}\exp \left( -j\frac{2\pi d_{\mathrm{d},k,q}}{\lambda} \right)\notag
				\\
				&\triangleq\exp \left( -j\frac{2\pi d_{\mathrm{d},k,q}}{\lambda} \right) \left( \sqrt{\frac{\kappa _{\mathrm{d}}\beta _{\mathrm{d},k,q}}{\kappa _{\mathrm{d}}+1}}+\sqrt{\frac{\kappa _{\mathrm{r}}\kappa _{\mathrm{t}}\beta _{\mathrm{r},k}\beta _{\mathrm{t},q}}{\left( \kappa _{\mathrm{r}}+1 \right) \left( \kappa _{\mathrm{t}}+1 \right)}}\sum_{m=1}^M{\exp \left( j\left( \psi _{k,m,q}+\theta _{m,q} \right) \right)} \right)
			\end{align}
			\hrulefill
		\end{figure*}
		\begin{figure*}%[hb]
			\begin{align}\label{a2}
				\left| a_{0,k,q} \right|^2
				&=\frac{\kappa _{\mathrm{d}}\beta _{\mathrm{d},k,q}}{\kappa _{\mathrm{d}}+1}+\frac{\kappa _{\mathrm{r}}\kappa _{\mathrm{t}}\beta _{\mathrm{r},k}\beta _{\mathrm{t},q}}{\left( \kappa _{\mathrm{r}}+1 \right) \left( \kappa _{\mathrm{t}}+1 \right)}\left| \sum_{m=1}^M{\exp \left( j\left( \psi _{k,m,q}+\theta _{m,q} \right) \right)} \right|^2 \notag
				\\
				&\quad +2\sqrt{\frac{\kappa _{\mathrm{d}}\kappa _{\mathrm{r}}\kappa _{\mathrm{t}}\beta _{\mathrm{d},k,q}\beta _{\mathrm{r},k}\beta _{\mathrm{t},q}}{\left( \kappa _{\mathrm{d}}+1 \right) \left( \kappa _{\mathrm{r}}+1 \right) \left( \kappa _{\mathrm{t}}+1 \right)}}\mathrm{Re}\left\{ \sum_{m=1}^M{\exp \left( j\left( \psi _{k,m,q}+\theta _{m,q} \right) \right)} \right\} \notag
				\\
				&=\frac{\kappa _{\mathrm{r}}\kappa _{\mathrm{t}}\beta _{\mathrm{r},k}\beta _{\mathrm{t},q}}{\left( \kappa _{\mathrm{r}}+1 \right) \left( \kappa _{\mathrm{t}}+1 \right)}{{\bm{\phi}} }_{q}^{\mathrm{H}}{{\bm \psi} }_{k,q}{{\bm \psi} }_{k,q}^{\mathrm{H}}{{\bm{\phi}} }_q+2\sqrt{\frac{\kappa _{\mathrm{d}}\kappa _{\mathrm{r}}\kappa _{\mathrm{t}}\beta _{\mathrm{d},k,q}\beta _{\mathrm{r},k}\beta _{\mathrm{t},q}}{\left( \kappa _{\mathrm{d}}+1 \right) \left( \kappa _{\mathrm{r}}+1 \right) \left( \kappa _{\mathrm{t}}+1 \right)}}\mathrm{Re}\left\{ {{\bm \psi} }_{k,q}^{\mathrm{H}}{{\bm{\phi}} }_q \right\} +\frac{\kappa _{\mathrm{d}}\beta _{\mathrm{d},k,q}}{\kappa _{\mathrm{d}}+1}
			\end{align}
			\hrulefill
		\end{figure*}
		Then, with the definitions in \eqref{varphi} and \eqref{Varphi}, $\left| a_{0,k,q} \right|^2$ is derived as shown in \eqref{a2} on the top of the next page.
		Finally, the expected received power $ \hat{P}_{k,q} =\mathbb{E} \left\{ P_{k,q} \right\}$ in \eqref{hatP_to_phi} can be obtained by using \eqref{P_k,l_origin} and \eqref{hatP_to_a}.
		This thus completes the proof of Theorem \ref{theorem_hatP}.
		
		\section{Proof of Theorem \ref{theorem_minorizing}}\label{appendix_minorizing}
		%		Recall that $ {\bm{\phi}} \triangleq \left[ {\bm{\phi}}_{1}^{\mathrm{T}},\dots ,{\bm{\phi}}_{L-1}^{\mathrm{T}} \right] ^{\mathrm{T}} $ and the constraints on $\left\{ {{\bm{\phi}} }_l \right\}$ is given by \eqref{constraint_theta_FHB}.
		Let $ {\mathcal R}_{\phi} =\left\{{\bm{\phi}} \mid \theta_{m,l}\in\left[ 0,2\pi \right) , \;m = 1, \dots,M, \;l \in \mathcal{L} _{\mathrm{b}}\right\} $ denote the feasible region of $ {\bm{\phi}} $.
		Based on the MM method, $ f\left( {\bm{\phi}} \right) $ is said to be minorized with $ \tilde{f}\left( {\bm{\phi}}\mid {\bm{\phi}}^r \right) $ if the following conditions are satisfied:
		\begin{enumerate}[({L}1)]
			\item $\tilde f\left( {{\bm{\phi}}\mid{{\bm{\phi}}^r}} \right)$ is continuous w.r.t. ${\bm{\phi}}$ and ${\bm{\phi}}^r$;
			
			\item $\tilde f\left( {{{\bm{\phi}}^r}\mid {{\bm{\phi}}}^r} \right) = f\left( {{{\bm{\phi}}^r}} \right),\forall {{\bm{\phi}}^r} \in {{\mathcal R}_{\phi}}$;
			
			\item $\tilde f\left( {{\bm{\phi}} \mid {\bm{\phi}}^r} \right) \leqslant f\left( {\bm{\phi}} \right),\forall {\bm{\phi}},{{\bm{\phi}}^r} \in {{\mathcal R}_{\phi}}$;
			
			\item $\tilde f'\left( {{\bm{\phi}} \mid {{\bm{\phi}}^r};{\bf{d}}} \right){\mid _{{\bm{\phi}} = {{\bm{\phi}}^r}}} = f'\left( {{{\bm{\phi}}^r};{\bf{d}}} \right)$ for $\forall {\bf{d}}$ belonging to the Boulingand tangent cone of ${\mathcal R}_{\phi}$.					
		\end{enumerate}
		Assume that the minorizing function $ \tilde{f}\left( {\bm{\phi}}\mid {\bm{\phi}}^r \right) $ has the following quadratic form
		\begin{align}\label{tildef_FHB_origin}
			\tilde{f}\left( {\bm{\phi}}\mid {\bm{\phi}}^r \right) &=f\left( {\bm{\phi}}^r \right) +2\mathrm{Re}\left\{ \mathbf{c}^{\mathrm{H}}\left( {\bm{\phi}}-{\bm{\phi}}^r \right) \right\}\notag \\
			&\quad+\left( {\bm{\phi}}-{\bm{\phi}}^r \right) ^{\mathrm{H}}\mathbf{C}\left( {\bm{\phi}}-{\bm{\phi}}^r \right),
		\end{align}
		where $\mathbf{c}\in \mathbb{C} ^{M\left( L-1 \right) \times 1}$ and $\mathbf{C}\in \mathbb{C} ^{M\left( L-1 \right) \times M\left( L-1 \right)}$ are undetermined parameters.
		Then,  (L1) and (L2) are always satisfied.
		In the following, we determine the expressions for $\mathbf{c}$ and $\mathbf{C}$ by (L3) and (L4).
		
		Let $ {\bm{\phi}}^{\mathrm{t}} $ be a feasible point in $ {{\mathcal R}_{\phi}} $.
		Then, the directional derivative of $\tilde{f}\left( {\bm{\phi}}\mid {\bm{\phi}}^r \right)$ at ${\bm{\phi}}^r$ with direction $\left( {\bm{\phi}}^{\mathrm{t}}-{\bm{\phi}}^r \right) $ is given by
		\begin{align}\label{direc_deriv_tildefphi}
			2\mathrm{Re}\left\{ \mathbf{c}^{\mathrm{H}}\left( {\bm{\phi}}^{\mathrm{t}}-{\bm{\phi}}^r \right) \right\}.
		\end{align}
		While the directional derivative of $ f\left( {\bm{\phi}} \right) $ is
		\begin{equation}\label{direc_deriv_fphi}
			2\mathrm{Re}\left\{ \left[ \sum_{k=1}^K{g_k\left( {\bm{\phi}}^r \right) \left( \mathbf{B}_{k}^{\mathrm{H}}{\bm{\phi}}^r+\mathbf{b}_k \right) ^{\mathrm{H}}} \right] \left( {\bm{\phi}}^{\mathrm{t}}-{\bm{\phi}}^r \right) \right\},
		\end{equation}
		where $ \mathbf{B}_k $, $ \mathbf{b}_k $ and $ g_k\left( {\bm{\phi}}^r \right) $ are defined in \eqref{B_k}, \eqref{b_k} and \eqref{g_k_phir}, respectively.
		From (L4), the directional derivatives \eqref{direc_deriv_tildefphi} and \eqref{direc_deriv_fphi} must be equal.
		By comparing coefficients, the expression of $\mathbf{c}$ can be identified as follows 
		\begin{equation}
			\mathbf{c}=\sum_{k=1}^K{g_k\left( {\bm{\phi}}^r \right) \left( \mathbf{B}_{k}^{\mathrm{H}}{\bm{\phi}}^r+\mathbf{b}_k \right)}.
		\end{equation}

		Denote by $\phi _{m,l}$ the $m$th element of ${\bm \phi} _l$.
		In the following, we consider a relaxed condition (L3') by replacing $ {\mathcal R}_{\phi} $ in (L3) with a relaxed feasible region $ {\mathcal R}_{\phi}^{\mathrm{relax}} \triangleq\left\{{\bm{\phi}} \mid \left| \phi _{m,l} \right|\leqslant 1 , \;m = 1, \dots,M, \;l \in \mathcal{L} _{\mathrm{b}}\right\} $.
		It will be shown that the conclusion also satisfies (L3).
		Specifically, we try to identify the expression of $\mathbf{C}$ by setting $\tilde{f}\left( {\bm{\phi}}\mid {\bm{\phi}}^r \right)$ as the lower bound of $ f\left( {\bm{\phi}} \right) $ 
		for each linear cut in any direction. 
		By introducing an auxiliary variable $\gamma \in \left[ 0,1 \right] $ and letting $ {\bm{\phi}}= {\bm{\phi}}^r+\gamma \left( {\bm{\phi}}^{\mathrm{t}}-{\bm{\phi}}^r \right) \in {\mathcal R}_{\phi}^{\mathrm{relax}}$, this sufficient condition is formulated as $ j\left( \gamma \right) \geqslant J\left( \gamma \right) $, where
		\newcounter{TempEqCnt} % 创建临时变量TempEqCnt
		\setcounter{TempEqCnt}{\value{equation}} % 将当前公式序号 赋给TempEqCnt
		\setcounter{equation}{97}
		\begin{figure*}%[hb]
			\begin{align}\label{lambda_min_lowerbound}
				\lambda _{\min}\left( {\bm \Lambda} \right) 
				&\overset{\left( \mathrm{a}1 \right)}{\geqslant}\sum_{k=1}^K{\tilde{g}_k\left( \gamma \right) \left( \lambda _{\min}\left( \left[ \begin{matrix}
						\mathbf{B}_k&		\mathbf{0}\\
						\mathbf{0}&		\mathbf{B}_{k}^{\mathrm{T}}\\
					\end{matrix} \right] \right) -\mu \lambda _{\max}\left( \left[ \begin{array}{c}
						\mathbf{e}_k\\
						\mathbf{e}_{k}^{*}\\
					\end{array} \right] \left[ \begin{array}{c}
						\mathbf{e}_k\\
						\mathbf{e}_{k}^{*}\\
					\end{array} \right] ^{\mathrm{H}} \right) \right)}+\mu \lambda _{\min}\left( \left[ \begin{array}{c}
					\sum\limits_{k=1}^K{\tilde{g}_k\left( \gamma \right) \mathbf{e}_k}\\
					\sum\limits_{k=1}^K{\tilde{g}_k\left( \gamma \right) \mathbf{e}_{k}^{*}}\\
				\end{array} \right] \left[ \begin{array}{c}
					\sum\limits_{k=1}^K{\tilde{g}_k\left( \gamma \right) \mathbf{e}_k}\\
					\sum\limits_{k=1}^K{\tilde{g}_k\left( \gamma \right) \mathbf{e}_{k}^{*}}\\
				\end{array} \right] ^{\mathrm{H}} \right) \notag
				\\
				&\overset{\left( \mathrm{a}2 \right)}{=}-2\mu \sum_{k=1}^K{\tilde{g}_k\left( \gamma \right) \mathbf{e}_{k}^{\mathrm{H}}\mathbf{e}_k}\overset{\left( \mathrm{a}3 \right)}{\geqslant}-2\mu \underset{k}{\max}\left\{ \mathbf{e}_{k}^{\mathrm{H}}\mathbf{e}_k \right\}
			\end{align}
			\hrulefill
		\end{figure*}
		\setcounter{equation}{99}
		\begin{figure*}%[hb]
			\begin{align}\label{e_k_square}
				\mathbf{e}_{k}^{\mathrm{H}}\mathbf{e}_k&=\left\| \mathbf{B}_{k}^{\mathrm{H}}\left( \gamma \left( {\bm{\phi}}^{\mathrm{t}}-{\bm{\phi}}^r \right) +{\bm{\phi}}^r \right) \right\| _{2}^{2}+\mathbf{b}_{k}^{\mathrm{H}}\mathbf{b}_k+2\mathrm{Re}\left\{ \mathbf{b}_{k}^{\mathrm{H}}\mathbf{B}_{k}^{\mathrm{H}}\left( \gamma \left( {\bm{\phi}}^{\mathrm{t}}-{\bm{\phi}}^r \right) +{\bm{\phi}}^r \right) \right\} \notag
				\\
				&\overset{\left( \mathrm{a}4 \right)}{\leqslant}\lambda _{\max}\left( \mathbf{B}_{k}^{\mathrm{H}}\mathbf{B}_k \right) \left\| \gamma \left( {\bm{\phi}}^{\mathrm{t}}-{\bm{\phi}}^r \right) +{\bm{\phi}}^r \right\| _{2}^{2}+\mathbf{b}_{k}^{\mathrm{H}}\mathbf{b}_k+2\mathrm{Re}\left\{ \mathbf{b}_{k}^{\mathrm{H}}\mathbf{B}_{k}^{\mathrm{H}}\left( \gamma \left( {\bm{\phi}}^{\mathrm{t}}-{\bm{\phi}}^r \right) +{\bm{\phi}}^r \right) \right\} \notag
				\\
				&\leqslant M\lambda _{\max}\left( \mathbf{B}_{k}^{\mathrm{H}}\mathbf{B}_k \right) +\mathbf{b}_{k}^{\mathrm{H}}\mathbf{b}_k+2\mathrm{Re}\left\{ \mathbf{b}_{k}^{\mathrm{H}}\mathbf{B}_{k}^{\mathrm{H}}\left( \gamma \left( {\bm{\phi}}^{\mathrm{t}}-{\bm{\phi}}^r \right) +{\bm{\phi}}^r \right) \right\} \notag
				\\
				&\overset{\left( \mathrm{a}5 \right)}{\leqslant}M\underset{l}{\max}\left\{ \lambda _{\max}\left( \left( \frac{\eta t_l}{E_{k}^{\mathrm{req}}} \right) ^2\mathbf{A}_{k,l}^{\mathrm{H}}\mathbf{A}_{k,l} \right) \right\} +\mathbf{b}_{k}^{\mathrm{H}}\mathbf{b}_k+2\left\| \mathbf{B}_k\mathbf{b}_k \right\| _1
			\end{align}
			\hrulefill
		\end{figure*}
		\setcounter{equation}{\value{TempEqCnt}} % 把TempEqCnt中存的公式序号赋回给当前公式序号
		\begin{align}
			j\left( \gamma \right) &\triangleq f\left( {\bm{\phi}}^r+\gamma \left( {\bm{\phi}}^{\mathrm{t}}-{\bm{\phi}}^r \right) \right),\label{j_define}
			\\
			J\left( \gamma \right) &\triangleq f\left( {\bm{\phi}}^r \right) +2\gamma \mathrm{Re}\left\{ \mathbf{c}^{\mathrm{H}}\left( {\bm{\phi}}^{\mathrm{t}}-{\bm{\phi}}^r \right) \right\} \notag \\
			&\quad+\gamma ^2\left( {\bm{\phi}}^{\mathrm{t}}-{\bm{\phi}}^r \right) ^{\mathrm{H}}\mathbf{C}\left( {\bm{\phi}}^{\mathrm{t}}-{\bm{\phi}}^r \right).\label{J_define}
		\end{align}
		It can be readily verified that $ j\left( 0 \right) =J\left( 0 \right) $.
		Denote by $ \nabla _{\gamma}j\left( \gamma \right) $ the first-order derivative of $j\left( \gamma \right) $ w.r.t. $ \gamma $.
		It can be derived that
		\begin{equation}
			\nabla _{\gamma}j\left( \gamma \right) =\sum_{k=1}^K{\tilde{g}_k\left( \gamma \right) \nabla _{\gamma}\tilde{h}_k\left( \gamma \right)},
		\end{equation}
		where 
		\begin{align}
			\tilde{h}_k\left( \gamma \right) &\triangleq h_k\left( {\bm{\phi}}^r+\gamma \left( {\bm{\phi}}^{\mathrm{t}}-{\bm{\phi}}^r \right) \right),
			\\
			\tilde{g}_k\left( \gamma \right) &\triangleq \frac{\exp \left(-\mu \tilde{h}_k\left( \gamma \right) \right)}{\sum_{k=1}^K{\exp \left( -\mu \tilde{h}_k\left( \gamma \right) \right)}}.
		\end{align}
		By defining $\mathbf{e}_k\triangleq \mathbf{B}_{k}^{\mathrm{H}}\left( \gamma \left( {\bm{\phi}}^{\mathrm{t}}-{\bm{\phi}}^r \right) +{\bm{\phi}}^r \right) +\mathbf{b}_k$ and $\tilde{{\bm{\phi}}}\triangleq {\bm{\phi}}^{\mathrm{t}}-{\bm{\phi}}^r$, we have $ \nabla _{\gamma}\tilde{h}_k\left( \gamma \right) =2\mathrm{Re}\left\{ \mathbf{e}_{k}^{\mathrm{H}}\tilde{{\bm{\phi}}} \right\} $.
		It is readily to verify that $\nabla _{\gamma}j\left( 0 \right) =\nabla _{\gamma}J\left( 0 \right) $.
		Therefore, a sufficient condition for $ j\left( \gamma \right) \geqslant J\left( \gamma \right) $ is 
		\begin{equation}
			\nabla _{\gamma}^{2}j\left( \gamma \right) \geqslant \nabla _{\gamma}^{2}J\left( \gamma \right) , \;\forall \gamma \in \left[ 0,1 \right].
		\end{equation}
		The second-order derivative of $ \tilde{h}_k\left( \gamma \right) $ is given by $ \nabla _{\gamma}^{2}\tilde{h}_k\left( \gamma \right)=2\tilde{{\bm{\phi}}}^{\mathrm{H}}\mathbf{B}_k\tilde{{\bm{\phi}}} $.
		Then, we have
		\begin{align}\label{j_second_derivate}
			\nabla _{\gamma}^{2}j\left( \gamma \right) &=\sum_{k=1}^K{\left( \tilde{g}_k\left( \gamma \right) \nabla _{\gamma}^{2}\tilde{h}_k\left( \gamma \right) -\mu \tilde{g}_k\left( \gamma \right) \left( \nabla _{\gamma}\tilde{h}_k\left( \gamma \right) \right) ^2 \right)}\notag \\
			&\quad+\mu \left( \sum_{k=1}^K{\tilde{g}_k\left( \gamma \right) \nabla _{\gamma}\tilde{h}_k\left( \gamma \right)} \right) ^2 \notag
			\\
			&=\left[ \begin{matrix}
				\tilde{{\bm{\phi}}}^{\mathrm{H}}&		\tilde{{\bm{\phi}}}^{\mathrm{T}}\\
			\end{matrix} \right] {\bm \Lambda} \left[ \begin{array}{c}
				\tilde{{\bm{\phi}}}\\
				\tilde{{\bm{\phi}}}^{\ast}\\
			\end{array} \right],
		\end{align}
		where
		\begin{align}
			{\bm \Lambda} &=\sum_{k=1}^K{\tilde{g}_k\left( \gamma \right) \left( \left[ \begin{matrix}
					\mathbf{B}_k&		\mathbf{0}\\
					\mathbf{0}&		\mathbf{B}_{k}^{\mathrm{T}}\\
				\end{matrix} \right] -\mu \left[ \begin{array}{c}
					\mathbf{e}_k\\
					\mathbf{e}_{k}^{*}\\
				\end{array} \right] \left[ \begin{array}{c}
					\mathbf{e}_k\\
					\mathbf{e}_{k}^{*}\\
				\end{array} \right] ^{\mathrm{H}} \right)}\notag \\
			&\quad+\mu \left[ \begin{array}{c}
				\sum_{k=1}^K{\tilde{g}_k\left( \gamma \right) \mathbf{e}_k}\\
				\sum_{k=1}^K{\tilde{g}_k\left( \gamma \right) \mathbf{e}_{k}^{*}}\\
			\end{array} \right] \left[ \begin{array}{c}
				\sum_{k=1}^K{\tilde{g}_k\left( \gamma \right) \mathbf{e}_k}\\
				\sum_{k=1}^K{\tilde{g}_k\left( \gamma \right) \mathbf{e}_{k}^{*}}\\
			\end{array} \right] ^{\mathrm{H}}.
		\end{align}
		In addition, the second-order derivative of $ J\left( \gamma \right) $ can be derived as 
		\begin{align}\label{J_second_derivate}
			\nabla _{\gamma}^{2}J\left( \gamma \right) 
			=2\tilde{{\bm{\phi}}}^{\mathrm{H}}\mathbf{C}\tilde{{\bm{\phi}}}
			=\left[ \begin{matrix}
				\tilde{{\bm{\phi}}}^{\mathrm{H}}&		\tilde{{\bm{\phi}}}^{\mathrm{T}}\\
			\end{matrix} \right] \left[ \begin{matrix}
				\mathbf{C}&		\mathbf{0}\\
				\mathbf{0}&		\mathbf{C}^{\mathrm{T}}\\
			\end{matrix} \right] \left[ \begin{array}{c}
				\tilde{{\bm{\phi}}}\\
				\tilde{{\bm{\phi}}}^{\ast}\\
			\end{array} \right].
		\end{align}
		From \eqref{j_second_derivate} and \eqref{J_second_derivate}, we have $ {\bm \Lambda} \succeq \left[ \begin{matrix}
			\mathbf{C}&		\mathbf{0}\\
			\mathbf{0}&		\mathbf{C}^{\mathrm{T}}\\
		\end{matrix} \right] $.
	
		Let us define a set of auxiliary matrices $ \left\{ \mathbf{\dot{B}}_{k,l}\right\}  $ that satisfies $\mathbf{B}_{k}=\sum \limits_{l \in \mathcal{L} _{\mathrm{b}}}{\mathbf{\dot{B}}_{k,l}}$ as follows
		\begin{equation}
			\mathbf{\dot{B}}_{k,l}=\left[ \begin{matrix}
				\mathbf{0}_{\left( l-1 \right) M}&		&		\\
				&		\mathbf{A}_{k,l}&		\\
				&		&		\mathbf{0}_{\left( L-l-1 \right) M}\\
			\end{matrix} \right].
		\end{equation}
		It can be readily verified that $\mathrm{rank}\left( \mathbf{\dot{B}}_{k,l} \right) =\mathrm{rank}\left( \mathbf{A}_{k,l} \right) =1$.
		Therefore, we have $\mathrm{rank}\left( \mathbf{B}_k \right) \leqslant \sum \limits_{l \in \mathcal{L} _{\mathrm{b}}}{\mathrm{rank}\left( \mathbf{\dot{B}}_{k,l} \right)}=L-1$ \cite[Eq. (1.6.4)]{Zhang2017Matrix}.
		Similarly, it can be proved that $ \left[ \begin{matrix}
			\mathbf{B}_k&		\mathbf{0}\\
			\mathbf{0}&		\mathbf{B}_{k}^{\mathrm{T}}\\
		\end{matrix} \right] $ is low rank, thus $ \lambda _{\min}\left( \left[ \begin{matrix}
		\mathbf{B}_k&		\mathbf{0}\\
		\mathbf{0}&		\mathbf{B}_{k}^{\mathrm{T}}\\
	\end{matrix} \right] \right)=0 $.
		To reduce the computational complexity of calculating $ \mathbf{C} $, we set $ \mathbf{C}=\alpha \mathbf{I} $, where 
		$ \alpha $ is a simple lower bound of $ \lambda _{\min}\left( {\bm \Lambda} \right) $.	
		To derive the expression of $ \alpha $, we first obtain the inequalities in \eqref{lambda_min_lowerbound} on the top of this page with the following lemmas:
		\setcounter{equation}{\value{equation}+1}
		\begin{enumerate}[({a}1)]
			\item ${\lambda _{\min }}\left( {\bf{A}} \right) + {\lambda _{\min }}\left( {\bf{B}} \right) \leqslant {\lambda _{\min }}\left( {{\bf{A}} + {\bf{B}}} \right)$, if ${\bf{A}}$ and ${\bf{B}}$ are Hermitian matrices \cite{lutkepohl1996handbook};
			
			\item ${\lambda _{\max }}\left( {\bf{A}} \right) = {\mathop{\rm Tr}} \left( {\bf{A}} \right)$ and ${\lambda _{\min }}\left( {\bf{A}} \right) = 0$, if ${\bf{A}}$ is rank one \cite{lutkepohl1996handbook};
			
			\item $\sum\nolimits_{m = 1}^M {{a_m}{b_m}}  \leqslant \max _{m = 1}^M\left\{ {{b_m}} \right\}$, if ${a_m},{b_m}\geqslant 0$ and $\sum\nolimits_{m = 1}^M {{a_m}}=1$ \cite[Theorem 30]{Matrix-differential};
			
			\item $\mathrm{Tr}\left[ \mathbf{AB} \right] \leqslant \lambda _{\max}\left( \mathbf{A} \right) \mathrm{Tr}\left[ \mathbf{B} \right] $, if ${\bf{A}}$ is positive semidifinite with maximum eigenvalue $ {\lambda _{\max }}\left( {\mathbf{A}} \right) $ and ${\bf{B}}$ is positive semidifinite \cite{lutkepohl1996handbook};
%			${\lambda _{\min }}\left( {\bf{A}} \right) + {\lambda _{\min }}\left( {\bf{B}} \right) \le {\lambda _{\min }}\left( {{\bf{A}} + {\bf{B}}} \right)$, if ${\bf{A}}$ and ${\bf{B}}$ are Hermitian matrices 
		\end{enumerate}
		
		In addition, it is readily to verify that
		\begin{enumerate}[({a}1)]\setcounter{enumi}{4}
			\item $\left[ \frac{\left[ \mathbf{B}_k\mathbf{b}_k \right] _1}{\left| \left[ \mathbf{B}_k\mathbf{b}_k \right] _1 \right|},\dots ,\frac{\left[ \mathbf{B}_k\mathbf{b}_k \right] _{ML}}{\left| \left[ \mathbf{B}_k\mathbf{b}_k \right] _{ML} \right|} \right] ^{\mathrm{T}}$ and $2\left\| \mathbf{B}_k\mathbf{b}_k \right\| _1$ are the optimal solution and optimal value of the following problem for $\mathbf{x}=\left[ x_1,\dots ,x_{ML} \right] ^{\mathrm{T}}$, respectively:
			\begin{subequations}
				\begin{alignat}{2}
					\max_{\bf{x}} \quad & 2\mathrm{Re}\left\{ \mathbf{b}_{k}^{\mathrm{H}}\mathbf{B}_{k}^{\mathrm{H}}\mathbf{x} \right\} \\
					\mbox{s.t.}\quad
					&\left| x_{ml} \right|\leqslant 1, \;\forall m=1,\cdots M, \;l=1,\cdots, L-1.
				\end{alignat}
			\end{subequations}
		\end{enumerate}
		Recall that $\mathbf{e}_k\triangleq \mathbf{B}_{k}^{\mathrm{H}}\left( \gamma \left( {\bm{\phi}}^{\mathrm{t}}-{\bm{\phi}}^r \right) +{\bm{\phi}}^r \right) +\mathbf{b}_k$.
		The term $ \mathbf{e}_{k}^{\mathrm{H}}\mathbf{e}_k $ can be transformed as shown in \eqref{e_k_square} on the top of this page.
		By substituting \eqref{e_k_square} into \eqref{lambda_min_lowerbound}, the expression of $\alpha$ in \eqref{alpha} is obtained.
		From (a5), it can be found that $ {\bm{\phi}}^r+\gamma \left( {\bm{\phi}}^{\mathrm{t}}-{\bm{\phi}}^r \right) \in {\mathcal R}_{\phi} $ when the last equality in \eqref{e_k_square} holds.
		In addition, the expression for $ \alpha $ is independent of $ \gamma $.
		Hence, the proof is completed.
%		This thus completes the proof of Theorem \ref{theorem_minorizing}.
		
	\end{appendices}

	\bibliographystyle{IEEEtran}
	\bibliography{IEEEabrv,Refer}

\end{document}